\documentclass[prd,amsmath,amsfonts,showpacs,preprintnumbers,%
               nofootinbib,showkeys,floatfix,twocolumn,a4paper]{revtex4}

\usepackage{amssymb,amsmath,amsfonts}
\usepackage{graphicx}
\usepackage{ifpdf}
\newcommand{\Eq}[1]{Eq.~(\ref{#1})}
\newcommand{\Eqs}[1]{Eqs.~(\ref{#1})}
\newcommand{\Fig}[1]{Fig.~\ref{#1}}
\newcommand{\Figs}[1]{Figs.~\ref{#1}}

\newcommand{\Sec}[1]{Sec.~\ref{#1}}
\newcommand{\App}[1]{Appendix~\ref{#1}}
\newcommand{\Tr}{\operatorname{tr}}                 % Tr
\newcommand{\ndf}{\mathsf{ndf}}                     % ndf
\newcommand{\identity}{\mathbf{1}}
\newcommand{\FP}{\mbox{F-P}}
\newcommand{\smallfrac}[2]{\mbox{\small ${\displaystyle
      \frac{#1}{#2}}$}}
\ifpdf
\graphicspath{{./Fig-PDF/}}
\else
\graphicspath{{./Fig/}}
\fi
\begin{document}

\preprint{ADP-08-12/T672}

 \title{Infrared exponents and the strong-coupling limit in lattice
 Landau gauge}
 \author{Andr\'e Sternbeck}
\thanks{Address since October 2009: Institut f\"ur Theoretische
  Physik, Universit\"at Regensburg, D-93040 Regensburg, Germany.}
 \author{Lorenz von~Smekal}
\thanks{Address since April 2009: Institut f\"ur Kernphysik, TU
  Darmstadt, Schlossgartenstr.~9, D-64289 Darmstadt, Germany.}
 \affiliation{
   Centre for the Subatomic Structure of Matter (CSSM), School of
   Chemistry \& Physics, The University of Adelaide, SA 5005,
   Australia
}

\date{April 9, 2010}

\begin{abstract}
  We study the gluon and ghost propagators of lattice Landau gauge in
  the strong-coupling limit $\beta = 0$ in pure $SU(2)$ lattice gauge
  theory to find evidence of the conformal infrared behavior of these
  propagators as predicted by a variety of functional continuum methods
  for asymptotically small momenta $q^2 \ll \Lambda_\mathrm{QCD}^2$. In
  the strong-coupling limit, this same behavior is obtained for the
  larger values of $a^2q^2$ (in units of the lattice spacing $a$), where
  it is otherwise swamped by the gauge field dynamics. Deviations for
  $a^2 q^2 < 1 $ are well parameterized by a transverse gluon mass
  $\propto 1/a$. Perhaps unexpectedly, these deviations are thus no
  finite-volume effect but persist in the infinite-volume limit. They
  furthermore depend on the definition of gauge fields on the lattice,
  while the asymptotic conformal behavior does not. We also comment on
  a misinterpretation of our results by Cucchieri and Mendes in 
  Phys.~Rev.~D81 (2010) 016005.
\end{abstract}



\maketitle

%---------------------------------------------------------------------------
\section{Introduction}
\label{sec:intro}

The Green's functions of QCD are the fundamental building blocks of
hadron phenomenology
\cite{Roberts:1994dr,Alkofer:2000wg,Fischer:2006ub,Roberts:2007jh}.
Moreover, their infrared behavior is also known to contain essential
information about the realization of confinement in the covariant
formulation of QCD, in terms of local quark and gluon field systems. 
In relation to the gluon and ghost propagators of Landau gauge QCD the
Dyson-Schwinger equation studies of
Refs.~\cite{vonSmekal:1997is,vonSmekal:1997vx} established that the
gluon propagator alone does not provide long-range interactions of a
strength sufficient to confine quarks, which dismissed a
widespread conjecture from the 1970's going back to the work of
Marciano, Pagels, Mandelstam and others.
It was concluded that the infrared dominant correlations are instead
mediated by the Faddeev-Popov ghosts of this formulation, whose
propagator was then found to be infrared enhanced, in agreement with
the Kugo-Ojima confinement criterion and thus consistent
with the conditions for confinement in local quantum field theory 
\cite{Alkofer:2000wg,Alkofer:2000mz,Lerche:2002ep}.

This infrared behavior was subsequently confirmed by a variety of
studies based on different functional methods in the continuum which
all led to the same conformal infrared behavior for the gluonic
 Green's functions of Landau gauge QCD. These include studies of their
Dyson-Schwinger Equations (DSEs) \cite{Lerche:2002ep}, Stochastic
Quantization \cite{Zwanziger:2001kw}, and of the Functional
Renormalization Group Equations (FRGEs)~\cite{Pawlowski:2003hq}.  
In fact, this conformal infrared behavior of gauge-field correlations
in non-Abelian gauge theories with confinement is directly tied to the
validity and applicability of the framework of local quantum field
theory to such theories.  

However, with the notable exception of pure $SU(2)$ lattice gauge theory
in two dimensions \cite{Maas:2007uv}, this is not what is being
observed with current lattice implementations of Landau gauge
\cite{Sternbeck:2006cg,Ilgenfritz:2006he,Sternbeck:2007ug,Cucchieri:2007md,
Bogolubsky:2007ud,Cucchieri:2007rg,Cucchieri:2008fc,Bogolubsky:2009dc}.
Rather, these are much more consistent with an essentially free
ghost propagator together with a massive and hence non-vanishing gluon
propagator in the infrared, in qualitative agreement with DSE
solutions proposed in the studies of
Refs.~\cite{Aguilar:2004sw,Boucaud:2006if,Dudal:2007cw,  
Aguilar:2008xm,Aguilar:2008fh,Boucaud:2008ji,Boucaud:2008ky}. 
Note that the functional equations of continuum quantum field theory
admit both types of solutions, the {\em scaling solution} with the
predicted conformal infrared behavior and the {\em decoupling
  solution} with a massive gluon propagator \cite{Fischer:2008uz}. At
present, the appealing aspects for fundamental reasons of the scaling
solution may be seen to stand against an overwhelming evidence for the
massive one, driven by the numerical results from present lattice
implementations of the Landau gauge. We will review the situation
briefly in the next section.   

In this paper we study the gluon and ghost propagators in the strong
coupling limit, $\beta \to 0$, of pure $SU(2)$ lattice Landau gauge
\cite{Sternbeck:2008wg,Sternbeck:2008na}. This limit can be
interpreted as the limit of infinite lattice spacing, $a \to \infty$,
at a fixed physical scale as set, {\it e.g.}, by the string tension
which behaves as $a^2\sigma \propto  - \ln\beta $ for 
$\beta\to0$.  Alternatively, when considering all momenta
in lattice units, $q \sim 1/a$ as we do here, the same
strong-coupling limit can also be interpreted as a hypothetic limit in
which all physical momentum scales such as the string tension,
$\Lambda_\mathrm{QCD}$ or the lowest glueball mass are sent to
infinity, {\it i.e.}, formally as $\sigma\to\infty$ or
$\Lambda_\mathrm{QCD}\to \infty$. Both interpretations are of course 
equivalent.  Either way, all momenta and masses in lattice units $1/a$
are infinitely small relative to the physical scale of the theory
which is precisely what we need for an analysis of the asymptotic
infrared behavior of its correlation functions.

By definition, there is then no scaling limit. All momenta are zero in
physical units. Having said that, we will nevertheless distinguish
large lattice momenta, with $a^2 q^2 \gg 1 $, from intermediate, $a^2
q^2 \sim 1$, and low momenta, $a^2 q^2 \ll 1$, in the strong-coupling
limit. Physically, they are all in the asymptotically-far infrared,
but for sufficiently large lattices, $L/a \gg 1$, only small momentum
modes will be affected by finite-size corrections which can be
assessed by varying the lattice size, $L/a$.

With this in mind, we will show that the conformal infrared behavior
is indeed observed at large lattice momenta in the strong-coupling
limit. Perhaps surprisingly, however, the deviations at intermediate
and small momenta reveal only a negligible (though systematic)
lattice-size dependence. In agreement with the standard lattice
Landau-gauge simulations at finite $\beta$, we find that the small
momentum behavior is well parameterized by a transverse gluon
mass. This is not predominantly due to the finite volume, but it
depends on the actual lattice discretization of the gauge fields,
while the large momentum-behavior does not. So the latter 
is consistent with scaling while the data at small
momenta shows decoupling, albeit being discretization dependent. 

The intermediate momentum region is characterized by the transition
between scaling and decoupling around $a^2 q^2 \approx 1$. This
transition sets a scale on an infinite lattice with $a/L\to 0$, which
corresponds to a transverse gluon mass  $a M $ of the order one in the
strong-coupling limit. This scale is thus not related
to the string tension or glueball masses in the
strong-coupling regime which all behave as
$a m \propto -\ln \beta $ for $\beta \to 0$. The occurance of such a
new infrared scale $aM \sim 1$ seems artificial and its discretization
dependence might in fact signal the breakdown of 
Slavnov-Taylor identities for minimal lattice Landau gauge 
in the strong-coupling regime.

The paper is organized as follows: In \Sec{sec:ISvD} we review the
expectations for the infrared behavior from the continuum studies as
background and motivation. The different parts of this section may be
consulted individually or skipped on a first reading as convenient.
In \Sec{sec:results4SLG} we show that our numerical results for the
Landau-gauge gluon and ghost propagators in the strong-coupling limit
do in fact show the scaling behavior for lattice momenta with $a^2 q^2
\gg 1$, that the critical exponent can be extracted in good agreement
with continuum predictions from this data, and that the deviations
from conformal scaling for $a^2 q^2 < 1 $ are well parameterized by a
transverse gluon mass $M \propto 1/a$ in the infinite-volume limit. 
In \Sec{sec:comparing_diff_latt_def} we compare various lattice
definitions of gauge potentials, which would all be equivalent in the continuum
limit, and show that the essential features of the scaling branch at
large lattice momenta, such as critical exponent and coupling, are unique.  
We furthermore demonstrate that the massive branch observed for
$a^2q^2 <1$ does depend on the lattice definition of the gluon fields,
and that it is thus not unambiguously defined. 
We interpret this as an ambiguity in the definition of Landau
gauge on the lattice which precludes a corresponding definition of 
a measure for gauge-orbit space in presence of Gribov copies
\cite{Gribov:1977wm}.  One might still hope that this 
ambiguity will go away at non-zero $\beta$ in the scaling limit. While
this is true at large momenta, we demonstrate in \Sec{sec:coupling}
that the ambiguity is still present in the low-momentum region, at least for
commonly used values of the lattice coupling such as $\beta = 2.3$ or
$\beta = 2.5$ in $SU(2)$. Our Summary and conclusions are provided in 
\Sec{sec:conclusion} and further technical details are given in 
two appendices.

\section{Infrared Scaling versus Decoupling}
\label{sec:ISvD}

The Landau-gauge gluon propagator, in (Euclidean) momentum space, is
parameterized by a single dressing function $Z$,
\begin{equation}
  \label{eq:gluonprop_dress}
  D^{ab}_{\mu\nu}(p) \,=\, 
  \delta^{ab}\left(\delta_{\mu\nu}-\frac{p_{\mu}p_{\nu}}{p^2}\right) 
  \frac{Z(p^2)}{p^2} \; ,
\end{equation}
and the ghost propagator by a corresponding dressing function $G$,
\begin{equation}
 \label{eq:ghostprop_dress}
  D_G^{ab}(p)   \,=\,  -\delta^{ab}\;\frac{G(p^2)}{p^2} \; .
\end{equation}
For their infrared behavior, {\it i.e.}, that of $Z(p^2)$ and $G(p^2)
$ for $p^2 \to 0$, we consider the two possibilities described in the
subsections that follow.

\subsection{Scaling}

The prediction of
\cite{vonSmekal:1997is,vonSmekal:1997vx,Alkofer:2000mz,Lerche:2002ep,
Zwanziger:2001kw,Pawlowski:2003hq}
amounts to infrared asymptotic forms
\begin{subequations}
   \label{eq:infrared-gh_gl}
\begin{align}
  \centering
  \label{eq:infrared-gl}
  Z(p^2) \, &\sim\, (p^2/\Lambda^2_\mathrm{QCD})^{2\kappa_Z} \; ,\\
 \label{eq:infrared-gh}
  G(p^2) \, &\sim \, 
  (p^2/\Lambda_\mathrm{QCD}^2)^{-\kappa_G} \; ,
\end{align}
\end{subequations}
for $p^2 \to 0$, which are both determined by a unique critical
infrared exponent
\begin{equation}
  \label{kappaZ=kappaG}
   \kappa_Z = \kappa_G \equiv \kappa \; , %\, \in (0.5,1) \;.
\end{equation}
with $ 0.5 < \kappa < 1$. Under a mild regularity assumption
on the ghost-gluon vertex \cite{Lerche:2002ep}, the value of this
exponent is furthermore obtained as \cite{Lerche:2002ep,Zwanziger:2001kw} 
\begin{equation} 
  \kappa \, = \, (93 - \sqrt{1201})/98 \, \approx \, 0.595 \; .
\label{kappa_c}
\end{equation}
The conformal nature of this infrared behavior in the pure Yang-Mills
sector of Landau gauge QCD is evident in the generalization to
arbitrary gluonic correlations \cite{Alkofer:2004it}: a uniform
infrared limit of one-particle irreducible vertex functions
$\Gamma^{m,n}$ with $m$ external gluon legs and $n$ pairs of
ghost/anti-ghost legs of the form
\begin{equation}
  \Gamma^{m,n} \, \sim\, (p^2/\Lambda_\mathrm{QCD}^2)^{(n-m)\kappa}\; ,   
\label{genIR}
\end{equation}
when all $p_i^2 \propto p^2 \to 0$, $i=1,\dots 2n+m$. In particular,
the ghost-gluon vertex is then infrared finite (with $n=m=1$) as it
must \cite{Taylor:1971ff}, and the nonperturbative running coupling 
introduced in \cite{vonSmekal:1997is,vonSmekal:1997vx} via the
definition
\begin{equation} 
 \alpha_s(p^2) \, = \, \frac{g^2}{4\pi} Z(p^2) G^2(p^2) 
\label{alpha_minimom}
\end{equation}
approaches an infrared fixed-point, $\alpha_s \to \alpha_c$ for $p^2 \to
0$. If the ghost-gluon vertex is regular at $p^2 =0$, its value is
\cite{Lerche:2002ep}
\begin{equation} 
\alpha_c \, = \, \frac{8\pi}{N_c} \, \frac{\Gamma^2(\kappa-1)
  \Gamma(4-2\kappa)}{\Gamma^2(-\kappa) \Gamma(2\kappa-1)} \,  \approx
%% \, \frac{2\pi}{N_c} \times \, 1.42  \; .  
  \, \frac{9}{N_c} \times \, 0.99  \; . 
  \label{eq:alphac}
\end{equation}
Comparing the infrared scaling behavior of DSE and FRGE solutions of
the form of Eqs.~(\ref{eq:infrared-gh_gl}), it has in fact been shown
that in presence of a single scale, the QCD scale
$\Lambda_\mathrm{QCD}$, the solution with the infrared behavior
(\ref{kappaZ=kappaG}) and (\ref{genIR}), with a positive exponent
$\kappa$, is unique \cite{Fischer:2006vf,Fischer:2009tn} and nowadays
being called the \emph{scaling solution}.

\subsection{Decoupling}

This uniqueness proof does not rule out, however, the possibility of a
solution with an infrared-finite gluon propagator, as arising from a
transverse gluon mass $M$, which then leads to an essentially free ghost
propagator, with the free massless-particle singularity at $p^2=0$,
{\it i.e.},
\begin{equation} 
  Z(p^2) \, \sim \, p^2/M^2\,, \quad \text{and} \quad G(p^2) \, \sim \,
  \mathrm{const.}
  \label{decoupling}
\end{equation}
for $p^2 \to 0$. The constant contribution to the zero-mo\-men\-tum gluon
propagator, $ D(0) = 1/M^2$, thereby necessarily leads to an
infrared constant ghost renormalization function $G$. This
solution corresponds to $\kappa_Z = 1/2 $ and $ \kappa_G = 0$. It does
not satisfy the scaling relations (\ref{kappaZ=kappaG}) or (\ref{genIR}).
This is because in this case the transverse gluons decouple
for momenta $p^2 \ll M^2 $, below the independent second scale given
by their mass $M$. It is thus not within the class of scaling
solutions considered above, and it is termed the {\em decoupling
  solution} in contradistinction. The interpretation of the
renormalization group invariant (\ref{alpha_minimom}) as a running
coupling does not make sense in the infrared in this case, in which there is
no infrared fixed-point and no conformal infrared behavior.  

\subsection{Continuum versus Lattice Studies}

The functional equations of continuum quantum field theory admit two
types of solutions. With the current implementations of Landau
gauge on the lattice strong support is provided for the massive
decoupling solution from simulations.  Is this the end of the story?
Superficially one might say that lattice gauge theory provides an ab
initio framework whereas functional continuum methods yield ambiguous
results. We would like to argue that this conclusion might be a bit
premature, however. First, lattice simulations must necessarily be
done in a finite volume where, strictly speaking, a conformal behavior
cannot be observed, certainly not for the lowest momentum values. What
is necessary to observe an at least approximate conformal behavior of
the correlation functions in a finite volume of extend $L$, is a wide
separation of scales,
\begin{equation}
  \pi/L \, \ll p \, \ll \Lambda_{\mathrm{QCD}} \; ,
\label{scales}
\end{equation}
such that a reasonably large number of modes with momenta $p$
sufficiently far below the QCD scale $\Lambda_\mathrm{QCD}$ are
accessible\footnote{The relevant scale here is that of
  the minimal MOM ($\mathsf{MM}$) scheme \cite{vonSmekal:2009ae}. Its
  relation to the $\mathsf{\overline{MS}}$ scheme, {\it
    e.g.}, for zero flavors and $N_c=3$,  with
  $\Lambda_{\overline{\mathsf{MS}}}\approx 240~\mathrm{MeV}$ 
  yields $\Lambda_{\mathsf{MM}} \approx450$~MeV. The enormous
  challenge for lattice simulations thus is to satisfy
  $\pi/L\ll p\ll 450$~MeV in the scaling region.}
whose corresponding wavelengths are at the same time much shorter
than the finite size $L$.\footnote{Finite-volume effects
have been analyzed carefully for the DSE scaling solution  
on a 4-dimensional torus which showed a quite significant volume
dependence \cite{Fischer:2007pf}. In particular, it was concluded that
volumes of about $15$~fm in length are needed before even an onset of
the leading infrared behavior can be observed, and that up to $40$~fm
might be required for a reliable quantitative determination of an
approximately conformal infrared behavior from a suitable range of
momenta satisfying (\ref{scales}). That this dependence does not
 match up with that on a lattice became clear at the 2007
Lattice Conference, where several results were reported
from large lattices
\cite{Sternbeck:2007ug,Bogolubsky:2007ud,Cucchieri:2007md}, the
largest for $SU(2)$ with up to 
$27$~fm in size \cite{Cucchieri:2007md}, showing 
practically no tendency to follow the predicted finite-size corrections to
the DSE scaling solution.} 
Secondly, gauge-fixing in
presence of Gribov copies is not so well understood on the lattice
either and the relation between present implementations of lattice
Landau gauge and covariant gauge theory in the continuum with local
Becchi-Rouet-Stora-Tyutin (BRST) symmetry is far from clear after all
\cite{vonSmekal:2008ws}. Last but not least, the functional continuum
methods are not as ambiguous as they might at first appear. At least
technically the fact they admit both kinds of solutions is not
surprising because it is well known that these methods have to be
supplemented by additional boundary conditions
\cite{Zwanziger:2001kw,Lerche:2002ep,Fischer:2008uz}. It is these boundary
conditions that determine which of the two solutions will be
obtained, and it is essentially these boundary conditions in which the
two classes of continuum studies mentioned above differ from one
another.  In particular, the scaling solution requires the boundary
condition for the subtraction in the ghost DSE such that  
\begin{equation} 
  G^{-1}(p^2) \, \to \, 0\,,\quad\text{for} \; p^2 \to 0\,,
\label{ir-dom-G}
\end{equation}
which in Landau gauge then implements both, the unbroken global gauge
charges of the Kugo-Ojima confinement criterion and the 
horizon condition of the original Gribov-Zwanziger framework
\cite{Gribov:1977wm,Zwanziger:1992qr}, by the infrared dominance of
ghosts. This implies that the Kugo-Ojima confinement criterion cannot be
derived from DSEs alone. It is implicitly implemented by the 
boundary condition (\ref{ir-dom-G}) which leads to the
conformal infrared behavior \cite{Lerche:2002ep}, while other choices genuinely
lead to the massive one \cite{Fischer:2008uz}.

More generally, from the functional equations alone (whether DSEs or
FRGEs or both together) both solutions are possible, in principle.
Because only the scaling solution is consistent with the conditions
for confinement in local quantum field theory, based on the cohomology
construction of a physical Hilbert space over the indefinite metric
spaces of covariant gauge theory from BRST symmetry, this appears to
be the physically relevant solution within this framework and it
therefore received a lot of attention in the functional continuum studies. 

The decoupling solution (\ref{decoupling}) has received renewed attention
\cite{Dudal:2007cw,Aguilar:2008xm,Aguilar:2008fh,Boucaud:2008ji,Boucaud:2008ky}
mainly because this is what is being observed in Landau gauge implementations
on the lattice 
\cite{Sternbeck:2006cg,Ilgenfritz:2006he,Sternbeck:2007ug,Cucchieri:2007md,
Bogolubsky:2007ud,Cucchieri:2007rg,Cucchieri:2008fc,Bogolubsky:2009dc}. 
The numerical procedures on the lattice are thereby based on
minimizations of a gauge-fixing potential with respect to gauge
transformations. To find absolute minima is not feasible on large
lattices as this is a non-polynomially hard computational problem. One
therefore settles for local minima which in one way or another,
depending on the algorithm, samples gauge copies of the first Gribov
region. This is mimicked in the continuum by the
inclusion of Zwanziger's horizon functional to suppress the gauge
copies outside the first Gribov region within the Gribov-Zwanziger
framework. To make this framework compatible with the decoupling
solution (\ref{decoupling}) one then introduces an additional mass
term \cite{Dudal:2007cw,Dudal:2008sp}. While the renormalizability is
maintained, one no-longer has an exact local BRST symmetry in this
framework which leads to unitarity violations when attempting a BRST
cohomology construction of a physical Hilbert space. This so-called
soft BRST breaking only matters in the non-perturbative regime and it
relates to the sampling of Gribov copies. A similar effect, in terms
of reweighting copies inside and outside the first Gribov horizon, 
is achieved by introducing an explicit Curci-Ferrari mass
\cite{vonSmekal:2008en}. In this case the reweighting is explicitly   
controlled by the mass and results in an explicit BRST breaking
proportional to that mass. It appears that a reweighting of Gribov
copies generally causes BRST breaking. At the moment it seems
questionable to us whether one can have a gluon mass without
reweighting of Gribov copies and BRST breaking.

Within the framework of local quantum field theory, which however
requires an unbroken BRST symmetry with nilpotent BRST charge, the
decoupling solution (\ref{decoupling}) is realized if and only if it
comes along with the Higgs mechanism. The Kugo-Ojima confinement
criterion and the infrared scaling of Landau gauge Green's functions
as a consequence of this criterion cannot be dismissed from lattice
simulations without a proper definition of a non-perturbative BRST
symmetry. It is worth remembering, however, that the apparent
ambiguity arises only when comparing the gauge dependent
Green's functions of either approach. By construction, physical
observables remain of course unaffected by this problem with BRST in
minimal lattice Landau gauge. One example is the Polyakov-loop
potential of the pure gauge theory whose center symmetry can be used
to define an alternative confinement criterion which is in fact
satisfied by the decoupling solution as well \cite{Braun:2007bx}, 
regardless of the realization of BRST symmetry.

The agreement between lattice Landau gauge and continuum (decoupling)
results, {\it i.e.}, when the restriction to the first Gribov region is
implemented, seems quite convincing. The fact that the accuracy and
conclusiveness of lattice results together with our understanding of
the functional methods based on local quantum field theory have
unveiled the conflict between the observed dynamical gluon mass and
BRST symmetry is a great achievement. We believe that it will eventually allow
us to understand the relation between Gribov copies and BRST symmetry.   

Meanwhile, the strong-coupling limit of lattice Landau gauge 
provides a powerful tool to study this ambiguity. In this
unphysical limit, the gluon and ghost correlations are solely driven
by what should correspond to a nonperturbative measure of gauge-orbit 
space in the Landau gauge, {\it i.e.}, the gauge-fixing and Faddeev-Popov parts
of the lattice measure. It is this measure that is being
assessed when the gauge-field dynamics is switched off. 

Ghost dominance, the essential condition for the
conformal infrared behavior, is then implemented by
hand, and if there is such a behavior, it should be seen at least in
this limit in which all momenta are asymptotically small (in physical
units). Because the strong-coupling limit can be interpreted as the
formal limit $\Lambda_\mathrm{QCD} \to \infty$, it is particularly
well suited to assess whether the conformal behavior of the scaling
solution is seen for the larger lattice momenta, with $a^2q^2\gg\pi^2
a^2/L^2$ and thus well clear of finite-size effects, after the upper
bound in (\ref{scales}) has been removed.  
This range of lattice momenta would otherwise be dominated by 
the dynamics due to the gauge action whose presence thus obscures 
any potentially conformal behavior there. As we demonstrate below,
its removal does in fact reveal a scaling behavior
(\ref{eq:infrared-gh_gl}), (\ref{kappaZ=kappaG}) for the first time in
lattice simulations, and it furthermore allows to study the
corrections to scaling from finite-volume effects on reasonably small
lattices with some systematics. 

It follows unambiguously from this study that finite-volume effects
play a minor role, however, and that up to these small effects, 
decoupling is observed at small lattice momenta $a^2 q^2 \ll 1$, with
a mass parameter $M \propto 1/a$. But the strong-coupling limit also 
serves to isolate a discretization ambiguity which manifests itself in
dependences on the lattice definition of gauge fields underlying the
respective lattice Landau gauges and their measures. 
While this noticeably affects the decoupling branch at $a^2
q^2 < 1$, the critical exponent and coupling of the scaling branch 
at large $a^2 q^2$ are rather insensitive to the discretization.%
\footnote{Note, however, that the extraction of a scaling exponent
  from the ghost propagator is affected by the Gribov ambiguity as 
  shown in the follow-up study of Ref.~\cite{Maas:2009ph}.}  

\begin{figure*}
  \centering
  \mbox{\includegraphics[height=7cm]{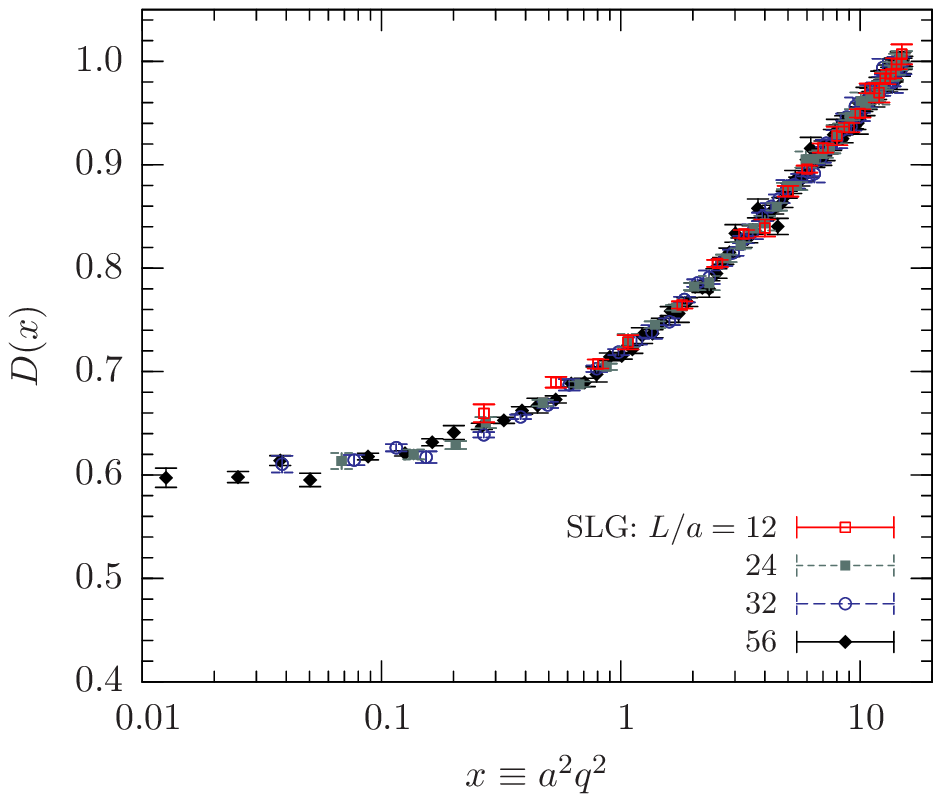}\qquad
    \includegraphics[height=7cm]{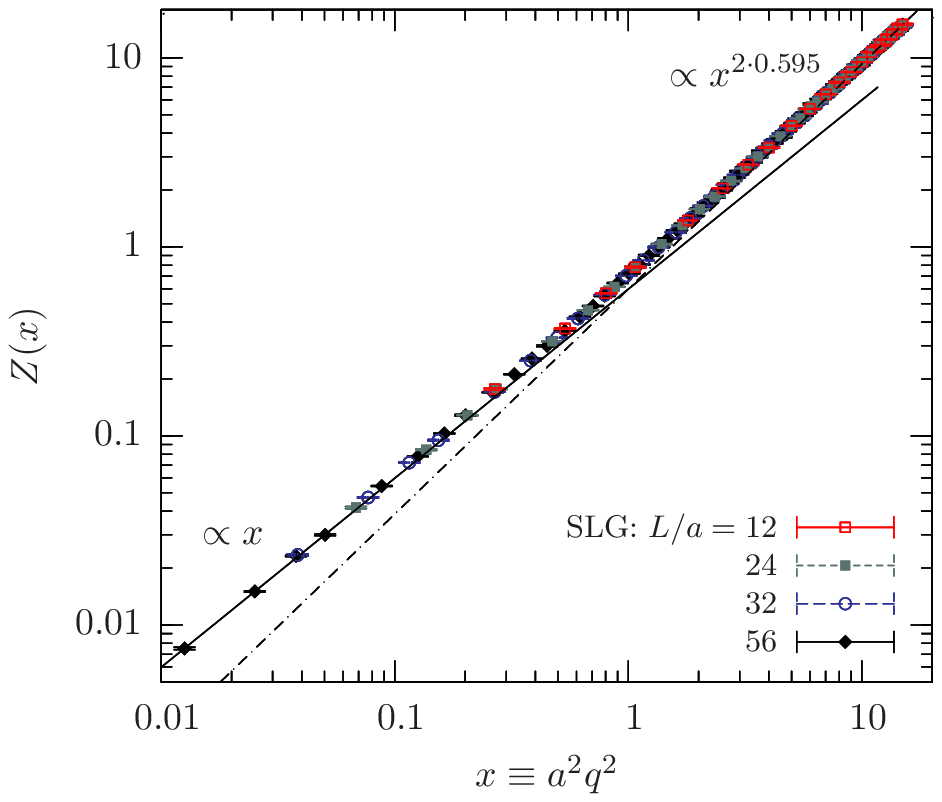}}
  \caption{The gluon propagator (left) and its dressing function $Z$
    (right) versus lattice momentum $x\equiv a^2q^2$ for different lattice
    sizes in the strong-coupling limit. For illustration purposes the
    data for $Z$ (right) is compared to the continuum predictions from
    decoupling (solid) and scaling (dashed) in infinite-volume limit.
    That is, the respective exponents have not been fitted but set to
    the expected decoupling and scaling values $\kappa=0.5$ and
    $\kappa\approx 0.595$.}
  \label{fig:gl_dress_qq_beta0-stdLG-U-Ud}  
\end{figure*}

%---------------------------------------------------------------------------
\section{Strong-coupling limit of standard lattice Landau gauge}
\label{sec:results4SLG}

We simulate pure $SU(2)$ gauge theory in the strong-coupling limit by
generating random link configurations $\{U\}$. These are sets of $SU(2)$
gauge links, 
\begin{equation}
  \label{eq:quaternion_repr}
  U_{x\mu} = u^0_{x\mu}\identity + i\sigma^a u^a_{x\mu}\,,
\end{equation}
equally distributed over~$(u^0,\vec{u})_{x\mu}\in S^3$. Those
configurations are then fixed to the standard lattice Landau gauge (SLG)
using an over-relaxation algorithm that iteratively minimizes the
$SU(2)$ gauge-fixing functional, for SLG,
\begin{equation}
  V_U[g] = 4 \sum_{x,\mu}\left(1-\smallfrac{1}{2} \Tr
    U^g_{x\mu}\right)
  \label{eq:functional}
\end{equation}
where the $U^g_{x\mu}=g_x U_{x\mu}g_{x+\hat\mu}^\dagger$ are the
gauge-transformed links. 
The Landau gauge condition for the stationarity of $V_U[g]$ under
gauge transformations $g$ is given by the lattice divergence, 
$F_x(A^g) = \nabla^b_\mu A_{x\mu}^g = 0$, where $\nabla^b_{\mu}$ denotes the
lattice backward derivative and $A_{x\mu}^g$ is the lattice gluon
field of SLG, defined by
\begin{equation}
  A_{x\mu}^g = \frac{1}{2ia}\left(U^g_{x\mu} -
    U^{g\,\dagger}_{x\mu}\right)
  \label{eq:gluonfield_SLG}
\end{equation}
in terms of the gauge-transformed link $U^g_{x\mu}$. 
To implement the minimal Landau gauge with a sufficient accuracy the
over-relaxation algorithm is iterated until the stopping criterion
\begin{equation}
  \label{eq:stoppingcriterion}
  \varepsilon:=\max_x\;\Tr\big[(\nabla^b_{\mu} A_{x\mu}^g)(\nabla^b_{\mu}
  {A^{g\,\dagger}_{x\mu}}) \big]< 10^{-13}
\end{equation}
is satisfied for every site on the lattice. Gluon and ghost
propagators are then calculated in momentum space employing standard
techniques. The dressing functions, $Z$ and $G$, are extracted
from the known tree-level form [\Eqs{eq:treelevelgluon} and
(\ref{eq:treelevelghost})] of the respective lattice propagator, see
\App{app:SLG} for further details.

\begin{figure}[b]
\centering
\includegraphics[width=\linewidth]{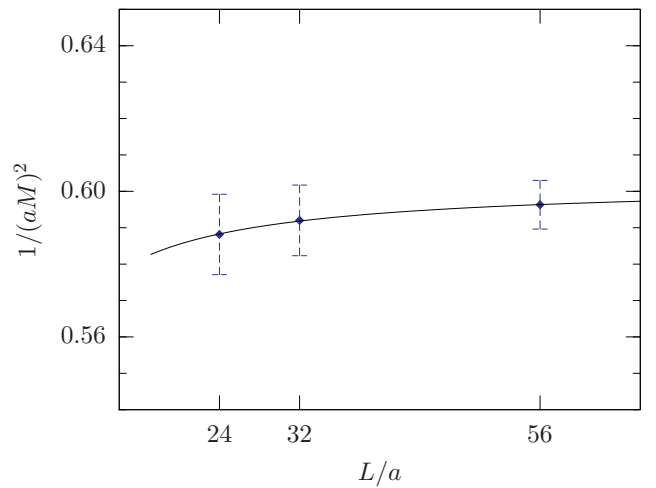}
\caption{The zero-momentum limit (\ref{eq:zeromom}) of the strong-coupling
  gluon propagator over $L/a$ from global fits 
  of the form (\ref{eq:gl_fit3}).} 
\label{fig:zeromom}
\end{figure}

\begin{figure*}
  \centering
  \mbox{\includegraphics[height=8.5cm]{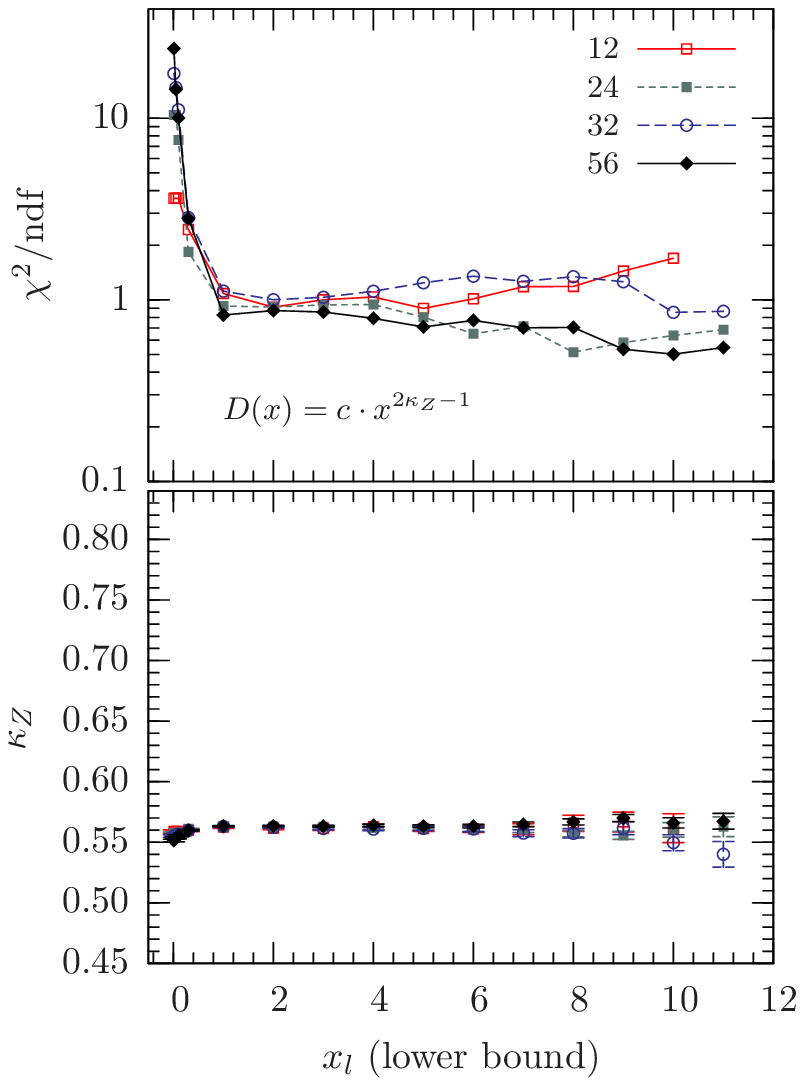}\hspace{-0.5cm}
  \includegraphics[height=8.5cm]{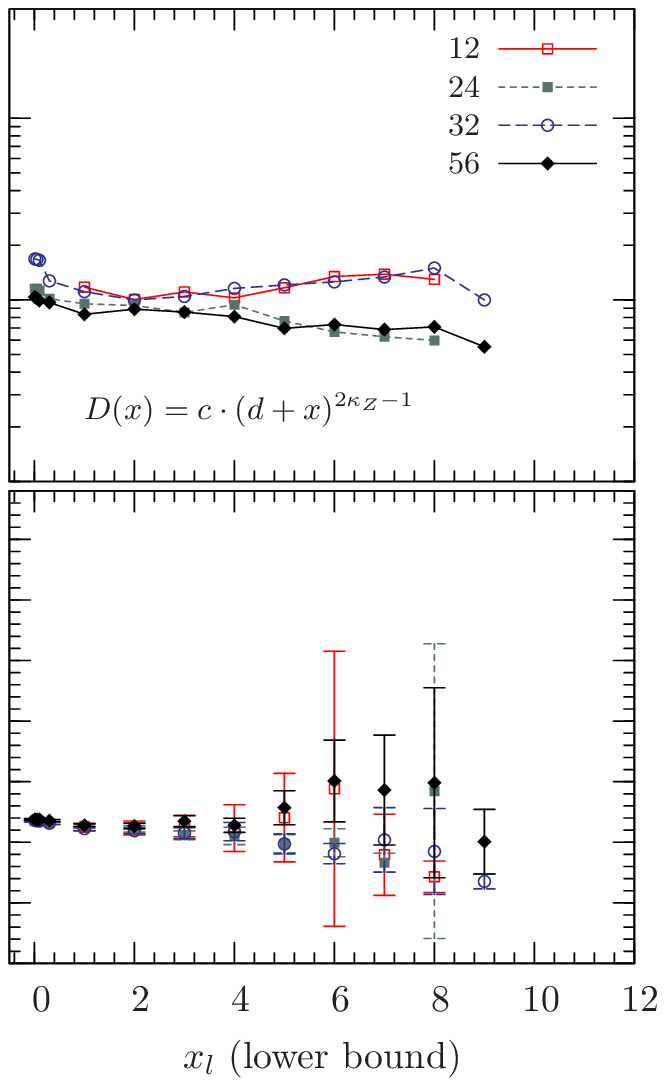}\hspace{-0.5cm}
  \includegraphics[height=8.5cm]{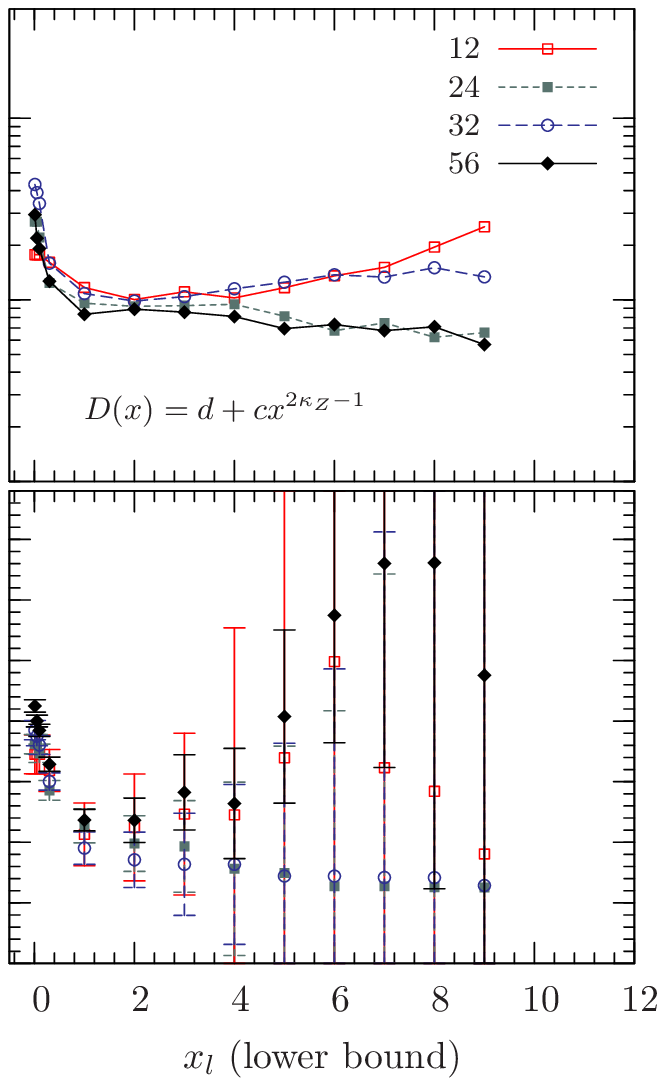}}
  \caption{$\chi^2/\ndf$ and $\kappa_Z$ values of fits to the gluon
    propagator data for different lattice sizes and different
    lower bounds $x_l$. The upper bound $x_u=14$ has been kept fixed. The
    respective fit functions are given in the upper panels.}
  \label{fig:kappa_gl_lowbound}
\end{figure*}

\subsection{Gluon propagator}

In \Fig{fig:gl_dress_qq_beta0-stdLG-U-Ud} the data for the gluon
propagator,  
\begin{equation} 
   D(x) \, \equiv \, Z(x)/x \; , \;\; 
     \mbox{with} \;\;  x\, \equiv \, a^2q^2\, ,
\end{equation}
and its dressing function $Z$ are
plotted against the lattice momenta $a^2q^2$ defined in
\Eq{eq:physical_momenta}.  
The propagator is observed to increase with momentum, while it 
plateaus at low momenta. Perhaps unexpectedly, however, this happens
irrespective of the lattice size ($N=L/a$) at around $a^2q^2 \approx
1$. It is therefore not primarily due to the finite volume. Rather, on
sufficiently large lattices the observed mass behaves as 
\begin{equation}
  M^2 \, \equiv \, \lim_{x\to 0}   D^{-1}(x) \, \propto \, 1/a^2
\label{eq:zeromom}
\end{equation}
in the strong-coupling limit with hardly any significant dependence on $L$.  
In particular, if there is a systematic $L$ dependence at all, 
the zero momentum limit of the gluon propagator tends to slowly
increase with the volume as shown in \Fig{fig:zeromom}. It certainly 
extrapolates to a finite value $\propto 1/a^2$ in the
limit $L/a \to \infty\,$.  

In the right panel of  \Fig{fig:gl_dress_qq_beta0-stdLG-U-Ud} we
furthermore compare the strong-coupling data for 
the gluon dressing function $Z$, 
to the predicted forms corresponding to decoupling (\ref{decoupling}),  
with $Z_{\mathrm{d}}(x) = c_{\mathrm{d}} x$, and scaling  
(\ref{eq:infrared-gl}), with $Z_{\mathrm{s}}(x) =
c_{\mathrm{s}}x^{2\kappa_Z}$ where the value of $\kappa_Z$ is not
fitted but taken from Eqs.~(\ref{kappaZ=kappaG}) and (\ref{kappa_c}),
$\kappa_Z = 0.595 $, for comparison.

With  $c_{\mathrm{d}}$ and $c_{\mathrm{s}}$  appropriately adjusted 
to the $56^4$ data, we find that the decoupling solution provides a
very good description of the low momentum region, while the large
momentum branch approaches the scaling solution with an exponent 
$\kappa_Z$ clearly above $0.5$ (the fitting procedure described below
leads to a conservative estimate of about  $\kappa_Z=0.57(1)$ for the
$56^4$ lattice, for example).   

In order to assess  the asymptotic form at large lattice  
momenta $x = a^2 q^2$ more quantitatively, we have fitted the gluon
propagator data to  the following three forms:
\begin{subequations}
 \label{eq:gl_ansatzes}
\begin{align}
  \label{eq:gl_fit1}
  D_a(x) &= cx^{2\kappa_Z-1}\,,\\
  \label{eq:gl_fit2}
  D_{b}(x) &= cx^{2\kappa_Z-1} + d\,,\\
  \label{eq:gl_fit3}
  D_{c}(x) &= c(d+x)^{2\kappa_Z-1}\,.
\end{align}
\end{subequations}
$D_a$ describes pure scaling with an effective exponent while $D_b$
and $D_c$ accommodate the transition between decoupling at small $x$
and scaling at large $x$ in different ways. In order to analyze the
scaling exponent $\kappa_Z$ we have used all three forms to fit the
data in various windows with increasing lower bound $x_l$. 
The results are fairly insensitive to variations of the upper bound
$x_u$ in some range sufficiently close to the maximum value of $x =
16$. We used $x_u = 14$ in all fits.

\begin{figure}[t]
\centering
\includegraphics[width=\linewidth]{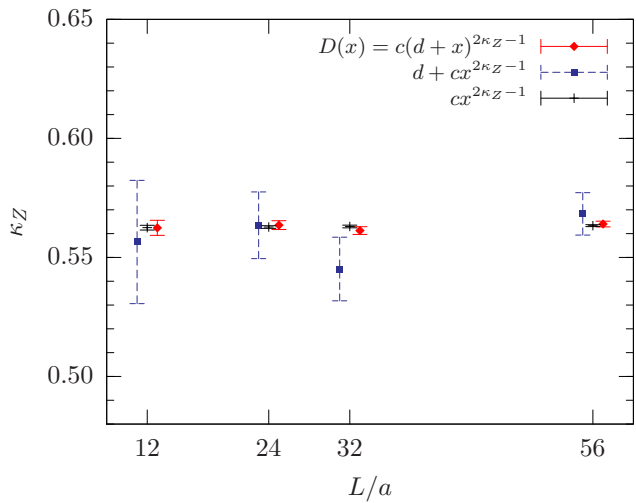}
\caption{The infrared exponent $\kappa_Z$ for lattice sizes
 $L/a=12,\, 24,\, 32$ and $56$ as obtained from the strong-coupling 
  gluon data on $x\in [1,14]$ with the three different fit models  in
 (\ref{eq:gl_ansatzes}).} 
\label{fig:kappa_gl_L_fits}
\end{figure}

The results of these fits for $\kappa_Z$ as functions of the lower
bound $x_l$ with the corresponding $\chi^2/\ndf$ are summarized in
\Fig{fig:kappa_gl_lowbound}. The pure scaling model $D_a(x)$ in
(\ref{eq:gl_fit1}) cannot describe the full momentum range but leads to 
good and stable fits to the data for $x_l \, {\scriptstyle \gtrsim} \,
1 $, the form  $D_c(x)$ in (\ref{eq:gl_fit3}) provides the best global
description of the data over the full momentum range. For $x_l$
between $1$ and $3$ the  results for $\kappa_Z$ from all three models are
consistent with each other within errors and all with
$\chi^2/\ndf$ of around $1$. The values of $\kappa_Z$ that result for the
different lattice sizes from all three fit models with $x_l =1$ are
shown in \Fig{fig:kappa_gl_L_fits}. 

As before, these values show very little systematic dependence on
$L/a$. A slight tendency to drift towards larger values in larger
volumes is observed, but this might not be a significant effect.
Assuming that there is no lattice size dependence to fit the values
from model $D_c(x)$ in (\ref{eq:gl_fit3}) for the 4 different lattice
sizes by a constant yields an average of $\kappa_Z = 0.563(1)$,
consistent with a global average over all values in
\Fig{fig:kappa_gl_L_fits}, while the $56^4$ data alone with the same
model gives $\kappa_Z= 0.564(1)$. For comparison, model $D_b(x)$ in
(\ref{eq:gl_fit2}), with the largest errors, for the $56^4 $ data on
$x\in [1,14]$ yields $ \kappa_Z= 0.568(9)$.

\subsection{Ghost dressing function}

\begin{figure}[t]
\centering
\includegraphics[width=\linewidth]{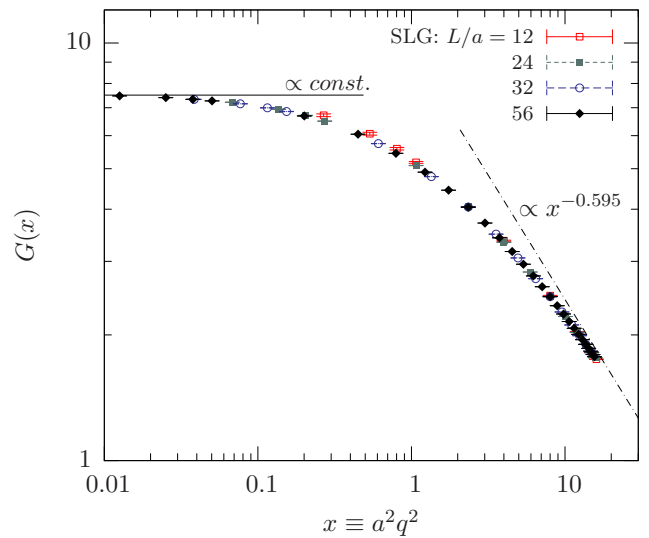}
\caption{The ghost dressing function $G$ for different lattice 
    sizes in the strong-coupling limit compared
    to the continuum predictions from decoupling (solid) and scaling
    (dashed) with the exponent from \Eq{kappa_c} 
    in infinite-volume limit (not fitted).}
\label{fig:gh_dress}
\end{figure}

We performed similar fits  to extract the exponent $\kappa_G$ from
the strong-coupling ghost dressing function $G$ as shown in
\Fig{fig:gh_dress}.  Those fits are less robust with a more pronounced
systematic uncertainty due to the fit model dependence. This 
is mainly because of the wider transition region, from $G = $ const.~at
small $x$ to $G\sim x^{-\kappa_G} $ at large $x$, which is under less
control here. In fact, a pure power law is at best observed only for
the very largest values of the lattice momentum, in the range above  
$x\approx 10$ or so. We again used three different fit models
analogous to those for the gluon data in (\ref{eq:gl_ansatzes}), one
describing pure scaling at large lattice momenta and two that interpolate
between decoupling (\ref{decoupling}) and scaling (\ref{eq:infrared-gh}). 
This time, however, we fit the inverse of the ghost dressing function $G$ 
as follows:
\begin{subequations}
 \label{eq:gh_ansatzes}
\begin{align}
  \label{eq:gh_fit1}
  G_a^{-1}(x) &= c x^{\kappa_G}\,,\\
  \label{eq:gh_fit2}
  G_{b}^{-1}(x) &= c x^{\kappa_G} + d\,,\\
  \label{eq:gh_fit3}
  G_{c}^{-1}(x) &= c (d+x)^{\kappa_G}\, .
\end{align}
\end{subequations}
With the same method as used above, the pure scaling form $G_a$ with
lower bounds $x_l$ around $12$ leads to values of $\kappa_G$ around 
$0.52$ for the $56^4$ lattice with a tendency to further 
increase with $x_l$, whereby the $\chi^2/\ndf \, $ decrease. But the
quality of those fits is still rather poor indicating that an
asymptotic scaling cannot be isolated from the transition region in
the ghost renormalization function. 

The fit models (\ref{eq:gh_fit2}) and (\ref{eq:gh_fit3}) take this
transition region better into account. The values for $\kappa_G$ from
$G_b$ in (\ref{eq:gh_fit2}) are the most stable ones but tend to be
rather large with $\kappa_G $ around $0.68$. Model $G_c$ leads to
exponents $\kappa_G $ most consistent with the scaling relation in
\Eq{kappaZ=kappaG}, with $\kappa_G$ in a range between $0.55$ and
$0.62$ depending on $x_l$.
\begin{figure}
  \includegraphics[width=\linewidth]{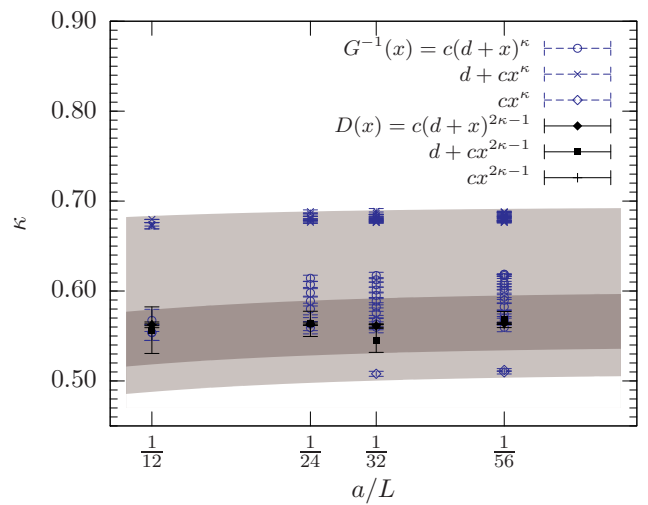}
  \caption{$\kappa$ versus $a/L$ for the ghost and gluon
    propagators. Grey-colored bands mark the variation of $\kappa$
    with the fit model and fitting window for either propagator.} 
  \label{fig:kappa_gl_gh_L_fits}
\end{figure}

The results for both exponents, $\kappa_Z$ and $\kappa_G$, are
summarized in \Fig{fig:kappa_gl_gh_L_fits}. For the gluon exponent
$\kappa_Z$ these are the same data points as in
\Fig{fig:kappa_gl_L_fits} with the dark grey band indicating the
errors and systematic uncertainties due to the fit model. The
$\kappa_G$ values are shown for a variety of lower bounds $x_l$ as
just described with model (\ref{eq:gh_fit3}) overlapping the gluon
results. The light grey band in \Fig{fig:kappa_gl_gh_L_fits} indicates
the considerably larger uncertainties in the ghost exponent, which are
mainly the systematic ones  due to the particular difficulty in
modelling the wide transition region between decoupling and scaling
there.  

Nevertheless, the set of all values extracted for $\kappa_G$ from the
$56^4$ data center around $\kappa_G = 0.60(8)$ which includes the
range for the gluon exponent and is thus fully consistent with the
scaling relation $\kappa_Z = \kappa_G$ in \Eq{kappaZ=kappaG}.

%---------------------------------------------------------------------------
\section{Different gauge-field definitions on the lattice}
\label{sec:comparing_diff_latt_def}

\begin{figure*}
  \begin{minipage}[t]{0.48\textwidth}
    \includegraphics[height=6.5cm]{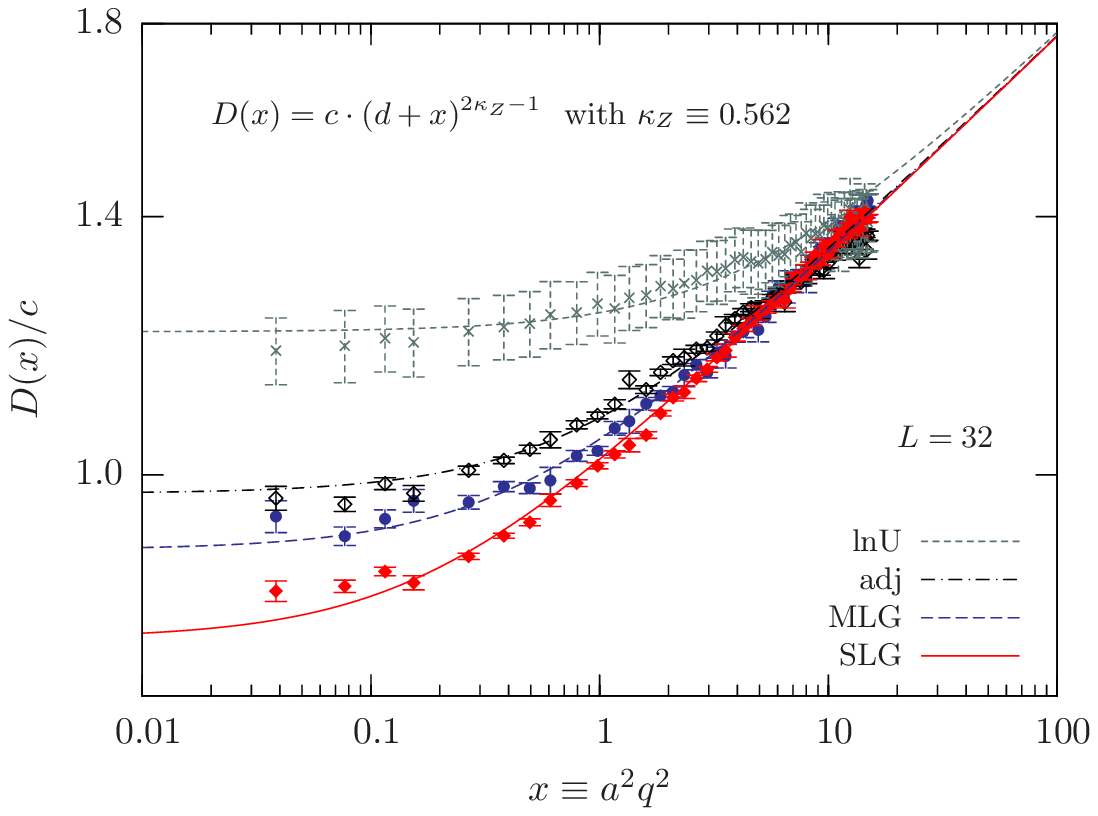}
    \caption{The strong-coupling gluon propagator over
      $a^2q^2$ for the various definitions of gauge fields in
      (\ref{compare}). All on $32^4$ lattices and normalized to the 
      scaling branch after fitting to $D_c$ in (\ref{eq:gl_fit3});
      all with $\kappa_Z=0.562$ from the fit to the SLG data.} 
    \label{fig:gl_qq_beta0}
  \end{minipage}
  \hfill
  \begin{minipage}[t]{0.48\textwidth}
      \includegraphics[height=6.5cm]{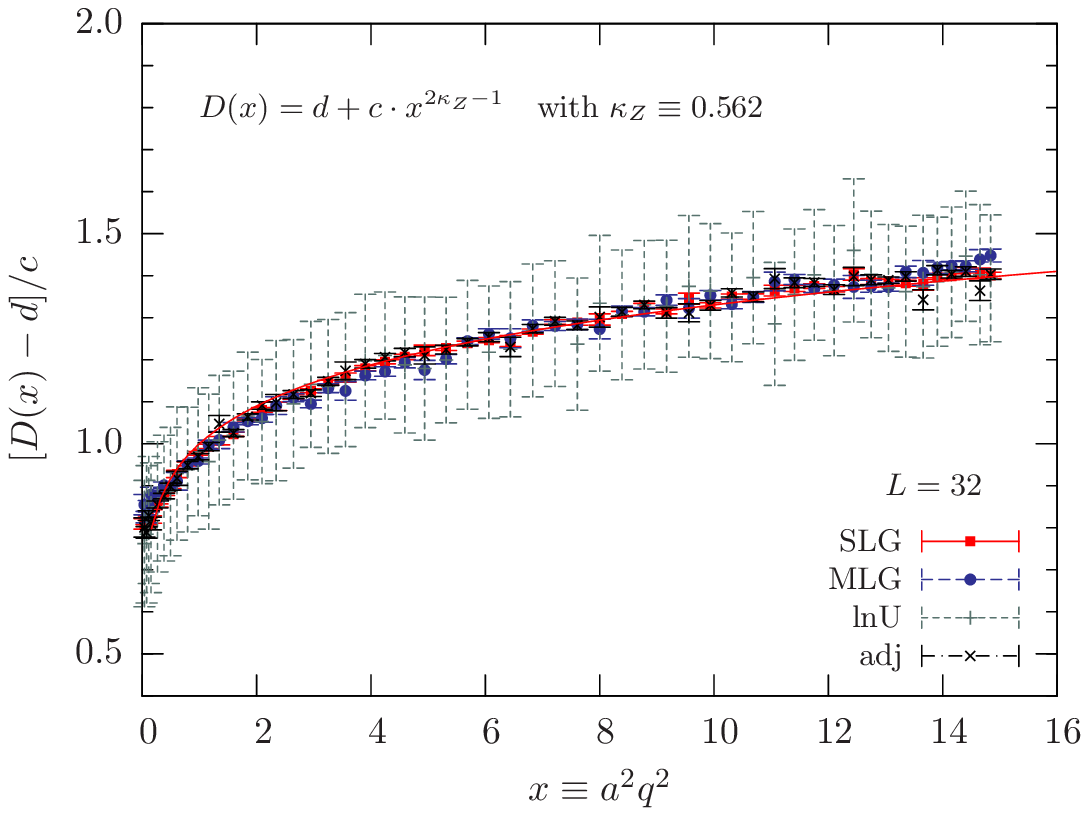}
    \caption{Same data as in \Fig{fig:gl_qq_beta0} but
      fitted to $D_b$ in (\ref{eq:gl_fit2}) with fixed $\kappa_Z=0.562$ and
      mass term subtracted in order to demonstrate how the data from
      all definitions then collapse onto a unique scaling curve $\propto
      x^{2\kappa_Z-1}$.}   
    \label{fig:global}
  \end{minipage}
\end{figure*}

Strong-coupling configurations are very rough with links
distributed uniformly over the parameter space, the 3-sphere for
$SU(2)$. The strong-coupling limit 
is therefore an ideal testbed for different lattice
definitions of gauge-fields which correspond to different choices of
coordinates that agree only near the identity, or in the continuum
limit. The standard definition (\ref{eq:gluonfield_SLG}) for example
corresponds to choosing separate coordinates for the Northern (NH) and
Southern Hemispheres (SH) of $S^3$ in the case of $SU(2)$. Strictly
speaking, the SLG gluon propagator therefore corresponds to an average
for each link of the contributions from NH and SH to the expectation
value in \Eq{eq:gluon}.    

The maximal chart is provided by stereographic projection which covers
the whole sphere except for the South Pole. A definition of $SU(2)$ gauge
fields on the lattice based on stereographic projection is possible as
follows,
\begin{equation} 
  \widetilde{A}_{x\mu} \,=\,  \frac{1}{2ia}\left(\widetilde{U}_{x\mu} -
    \widetilde{U}^{\dagger}_{x\mu}\right)\; , 
  \label{eq:gluonfield_stereo_1}
\end{equation}
where
\begin{equation}
  \widetilde{U}_{x\mu} 
     \, \equiv\,  \frac{2}{1+\frac{1}{2} \Tr U_{x\mu}} \;
     U_{x\mu} \;.
  \label{eq:gluonfield_stereo_2}
\end{equation}
It agrees with the standard definition near the North Pole, and in the
continuum limit, but the South Pole is now at infinity and the gauge
fields in (\ref{eq:gluonfield_stereo_1}) are thus non-compact variables. 
The associated Landau gauge is the \emph{modified lattice Landau
  gauge} (MLG) of Ref.~\cite{vonSmekal:2007ns}. It follows from the
stationarity condition of the modified gauge-fixing functional,  
\begin{equation}
  \label{eq:functional_modLG}
  \widetilde V_U[g] = -8 \sum_{x,\mu}
  \ln\left(\smallfrac{1}{2}+\smallfrac{1}{4}\Tr 
    U^{g}_{x\mu} \right)\; , 
\end{equation}
with respect to gauge transformations $g$, and reads 
\begin{equation} 
 \widetilde F_x(A^g) \, = \, F(\widetilde A^g) \,= \,
 \nabla^b_\mu \widetilde A^g_{x\mu} \, = \, 0\;.  
\end{equation}
When comparing MLG to the ever popular SLG, there is no advantage that the
SLG has over the MLG. A promising particular feature of the MLG on the
other hand is that it provides a way to perform gauge-fixed MC
simulations sampling \emph{all} Gribov copies of either sign (of the
Faddeev-Popov determinant) in the spirit of BRST. This feature will be
explored in a forthcoming study.  Here we simply use the MLG for
comparison in the standard way, {\it i.e.}, we gauge-fix configurations via
minimization of the MLG functional in \Eq{eq:functional_modLG}. 
The \FP~operator of the MLG is given in \eqref{mod_M_FP} in \App{app:MLG}.

Both lattice definitions of
Landau gauge have the same continuum limit, and any differences
between MLG and SLG data at finite lattice spacings are lattice
artifacts. It is also worth mentioning that gauge configurations fixed
to MLG do not satisfy the gauge condition of SLG and vice
versa. Nevertheless, exact transversality, {\it i.e.},\ $q_{\mu}(k)
A_{\mu}(k) = 0$ or $q_{\mu}(k)
\widetilde A_{\mu}(k) = 0$, is satisfied at finite lattice spacing 
  $a$ for both of them equally, with their respective Lorenz
conditions $\nabla^b_\mu A_{x\mu} = 0 $ or $\nabla^b_\mu \widetilde
A_{x\mu} = 0 $ and midpoint definition, if the momenta $q_\mu(k)$ are
defined as  ~$aq_{\mu}(k) = 2\sin(\pi k_{\mu}/N_{\mu})$ with integer valued
$k_{\mu}\in(-N_{\mu}/2,N_{\mu}/2]$. The identification of this so
  defined $q_\mu(k) $ with physical momentum is the usual tree-level
correction for the Wilson gauge action. 
It is a special feature of MLG and SLG that their lattice Landau gauge
conditions define gluon fields that are transverse in this physical
momentum at any finite lattice spacing. 

\subsection{Gluon propagator}

The data for the gluon propagator of SLG (red filled diamonds) is
compared to that of MLG (blue filled circles) in
\Fig{fig:gl_qq_beta0}. There we also show data for the gluon
propagator where either
\begin{align}
  \label{eq:adjdef}
    a A^{\mathsf{adj}}_{x\mu} &=\, u^0_{x\mu} u^{a}_{x\mu}\sigma^a
    \quad\mbox{(no sum $\mu$)}\;,   
\intertext{or}
  \label{eq:lnUdef}
    a A^{\mathsf{ln}}_{x\mu} &=\,
    \phi^a_{x\mu}\sigma^a/2\quad\text{from}\quad
    U_{x\mu}=\, e^{i \phi^a_{x\mu}\sigma^a/2}
\end{align}
were used to define lattice gluon fields based on the adjoint
representation, $A^{\mathsf{adj}}$ (black open diamonds), and thus
blind to the center \cite{Langfeld:2001cz}; or on the tangent space at
the identity $A^{\mathsf{ln}}$ (green crosses).  In these two cases,
$A^{\mathsf{adj}}$ and $A^{\mathsf{ln}}$, for the purpose of a
qualitative comparison, we simply use the gauge configurations of the
SLG to calculate the gluon propagator. Especially for
$A^{\mathsf{ln}}$ this implies, however, that the condition
$q_{\mu}(k)A_{\mu}(k)=0$ is satisfied at best approximately and
nowhere near the precision of the other two (SLG and MLG). The
residual uncertainty due to other possible tensor structures in the
gluon propagator (\ref{eq:gluon}) then causes the somewhat larger
errors for this definition as seen in \Figs{fig:gl_qq_beta0} and
\ref{fig:global}.

In \Fig{fig:gl_qq_beta0} we have first fitted the data from all four
definitions to $D_{c}$ in \Eq{eq:gl_fit3} which provides the best
overall description in the full momentum range as mentioned above. In
order to demonstrate how the other definitions compare to the SLG, we
keep its value for the exponent fixed when fitting the other data,
{\it i.e.}, $\kappa_Z = 0.562$ as obtained for $L/a = 32$ in SLG is
used in all fits. Relative to the scaling branch $\propto
x^{2\kappa_Z-1}$ for large $x\equiv a^2q^2 $ we then observe a strong
definition dependence in the (transverse) gluon mass term at
small~$x$. The relative weight of the two asymptotic branches, scaling
at large $x$ and massive at small, is clearly discretization
dependent, and this dependence cannot be compensated by finite
renormalizations as is manifest in the data of \Fig{fig:gl_qq_beta0}.

The fact that the observed mass from the zero-momentum limit
(\ref{eq:zeromom}) behaves as $M\propto 1/a$ is a first indication
that it is indeed this massive branch which is the ambiguous one. This is 
consistent with the fact that the definitions of gauge fields on the
lattice, which agree at leading order, all differ at order $a^2$. In
particular, with 
\begin{equation}
   U \,= \, \cos\smallfrac{\phi}{2} + i \vec\sigma \cdot
   \hat{\vec\phi} \, \sin\smallfrac{\phi}{2}  \quad \mbox{and} 
   \;\; \vec A \, = \Tr \vec\sigma A\; ,   
\end{equation}
for $SU(2)$, the four different gauge field definitions
(\ref{eq:gluonfield_SLG}), (\ref{eq:gluonfield_stereo_1}),
(\ref{eq:adjdef}) and (\ref{eq:lnUdef}) simply correspond to
\begin{subequations}
\label{compare}
\begin{align} 
 a \vec A &=\, \hat{\vec\phi} \;  {2} \sin\smallfrac{\phi}{2}\\
  a \vec{\widetilde A} &=\, \hat{\vec\phi} \;  {4}
  \tan\smallfrac{\phi}{4}\\
  a \vec A^{\mathsf{adj}} &=\, \hat{\vec\phi} \; \sin\phi \\
 a \vec A^{\mathsf{ln}} &=\, \vec\phi \; , \quad \mbox{with} \; \;
 \vec\phi = \hat{\vec\phi}\, \phi\;, \; \; \hat{\vec\phi}^2 = 1 \; ,
\end{align}
\end{subequations}
which clearly all agree only at leading order in the limit $a\to 0$.
From the order $a^2$ differences, the corresponding Jacobian factors
lead to likewise different lattice mass counter-terms for each of the
4 definitions. This is well known from lattice
perturbation theory where the lattice Slavnov-Taylor
identities guarantee, however, that the gluon remains massless at every order
by cancellation of all quadratically divergent contributions to its
self-energy for each of the definitions. In the strong-coupling limit
of minimal lattice Landau gauge, with an effective decoupling mass
(\ref{decoupling}) behaving as $M^2 \propto 1/a^2$, such a contribution
survives. This contribution depends on the measure for gauge
fields whose definition from minimal lattice Landau gauge beyond
perturbation theory is therefore ambiguous. The observation that these
differences matter here explicitly demonstrates the breakdown of the
lattice Slavnov-Taylor identities in minimal lattice Landau gauge in
the non-perturbative domain.

To assess whether this ambiguity has an influence on the exponent
$\kappa_Z$, we have also used the fits to the form 
$D_{b}$ in \Eq{eq:gl_fit2}, again with $\kappa_Z\equiv0.562$ fixed
from the $32^4$ SLG data. This fit model leads to somewhat larger
$\chi^2/\ndf$ arising from the transition region around $a^2q^2 \sim
1$ which this form does not describe quite as well as $D_{c}$, see the
discussion in the previous section. 
Having obtained the fits to $D_b$ allows us to subtract the
constant~$d$, however, which then makes the normalized data of
all four definitions nicely collapse onto a unique curve $\propto 
x^{2\kappa_Z-1}$ as seen in \Fig{fig:global}. 

The scaling region in the strong-coupling data for the gluon
propagator from all 4 definitions is fully consistent with a unique
exponent $\kappa_Z $ of around the SLG value with a conservative estimate
of an infinite-volume extrapolation of $\kappa_Z = 0.57(3)$.

\subsection{Ghost dressing function}

\begin{figure}[t]
  \centering
  \includegraphics[width=\linewidth]{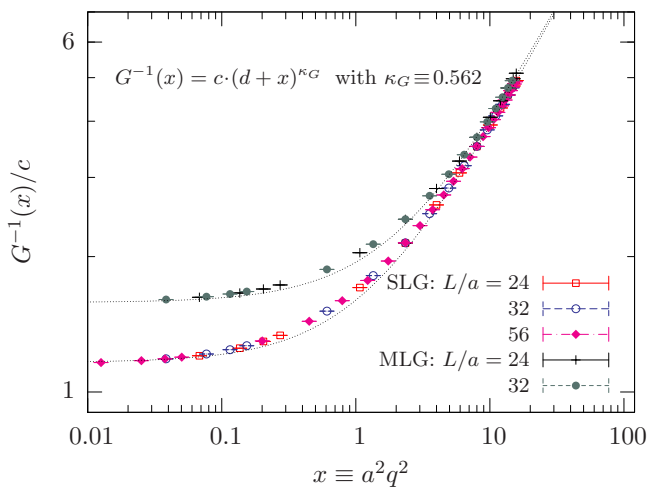}  
  \caption{Inverse ghost dressing functions in the strong-coupling
    limit of minimal lattice Landau gauge using standard (SLG) and
    modified (MLG) gauge fields/conditions.}
  \label{fig:ghinv_dress_qq_beta0}
\end{figure}

Equally consistent definitions of transverse gauge-fields, gauge conditions
and Faddeev-Popov operators are available for SLG and MLG which 
allow us to compare their respective ghost propagators also. In 
\Fig{fig:ghinv_dress_qq_beta0} we show the strong-coupling data for
the (inverse) ghost dressing functions from these two alternative
definitions of lattice Landau gauge, again normalized to the scaling
branch at large momenta. For this normalization we have fitted the
data to model $G_c$ in (\ref{eq:gh_fit3}) assuming 
the scaling relation $\kappa_G = \kappa_Z$ with the value of
$\kappa_Z = 0.562$ from the $32^4 $ SLG gluon data.  
We then again observe that the two definitions approach each other in
the scaling branch $\propto x^{-\kappa}$ at large $x$ but deviate in
the decoupling branch at small. Within the fit model uncertainties the
ghost data from both definitions is again consistent with a unique scaling
exponent, and with the scaling relation $\kappa_G = \kappa_Z$. 
And again the deviations in relative strengths of the two branches
cannot be compensated by renormalization. 

Note that in both cases, for the strong-coupling gluon and ghost
propagators, the normalization constants determined from their
respective scaling branches are actually not arbitrary. Their product
is related to the critical coupling (\ref{eq:alphac})
and thus not independent but also unique as we will discuss next. This
is consistent with the conclusion that it is the decoupling behavior
at low momenta which is ambiguous but not the scaling behavior at large.

%-----------------------------------------------------------------------------
\section{Running coupling}
\label{sec:coupling}

The renormalization-group invariant product of Landau gauge gluon and
ghost dressing functions (\ref{alpha_minimom}) defines a running
coupling.  Its perturbative behavior can be determined from
unrenormalized bare lattice data for the minimal Landau gauge
propagators
\cite{Sternbeck:2007br,Boucaud:2008gn,Sternbeck:2010xu}. Its relation
to the running coupling in the $\overline{\mathsf{MS}}$ scheme is
known to four loops and it can provide a valuable alternative to the
$\overline{\mathsf{MS}}$ coupling in phenomenological applications
\cite{vonSmekal:2009ae}. %,vonSmekal:2008ma,Sternbeck:2008au}.  
This furthermore permits an independent additional lattice determination of
the QCD scale parameter $\Lambda_{\overline{\mathsf{MS}}}\,$ from
continuum extrapolation of the bare product of the lattice Landau
gauge propagators
\cite{Sternbeck:2007br,Boucaud:2008gn,Sternbeck:2010xu}.

Without renormalization the bare lattice propagators are normalized
so as to reproduce their respective  tree-level forms
[Eqs.~\eqref{eq:treelevelgluon} and \eqref{eq:treelevelghost}] for the trivial
link configuration, when all links $U$ are set to the identity element.
Then, however, there is no significance in the constant prefactors of
the individual propagators. This is why we removed these overall
constants when comparing SLG and MLG data for these propagators
individually as in \Figs{fig:gl_qq_beta0} and
\ref{fig:ghinv_dress_qq_beta0} above. Their product
(\ref{alpha_minimom}) does not get renormalized, however, and should
therefore be independent of the lattice definition used for the gauge
fields up to discretization errors. We will discuss this definition
independence separately in the strong-coupling limit, where the  discretization
effects are largest, and at finite lattice couplings $\beta$ where it
provides an important consistency check for the lattice determinations
of the QCD scale parameter from the large momentum data of this strong
running coupling as described above and in more detail in 
Ref.~\cite{Sternbeck:2010xu}.

%-----------------------------------------------------------------------------
\subsection{Strong-coupling limit}
\label{sec:couplingconstant}

The predicted infrared scaling (\ref{eq:infrared-gh_gl}) with
$\kappa_Z = \kappa_G$ immediately implies that the running coupling
defined in \Eq{alpha_minimom} approaches an infrared fixed point,
$\alpha_s \to \alpha_c $ for $p^2 
\to 0 $. Standard continuum conventions of course need rescaling
$g^2 Z \to Z$  when comparing to lattice definitions such as
(\ref{eq:gluonfield_SLG}). The predicted conformal scaling in the
strong-coupling limit, with 
\begin{equation} 
 Z \,=\, c_Z \, (a^2q^2)^{2\kappa} \quad \mbox{and}  \quad G^{-1} \, =\,
 c_G\, (a^2q^2)^\kappa , 
\end{equation}
would therefore imply that the coupling 
(\ref{alpha_minimom}) should be constant with  
\begin{equation}
   \alpha_s \, =\,  \alpha_c \, =\,  c_Z/(4\pi c_G^2)\; .
\end{equation}
Note that its value is thus determined precisely by those
multiplicative constants in the propagators that were irrelevant to
the analysis in the previous section. They have to be 
extracted from the bare lattice data without rescaling or
renormalization. Besides the critical scaling exponent $\kappa $,
the critical coupling $\alpha_c$ is  an independent additional
prediction from infrared scaling and it is determined by
these constants.

\begin{figure}
  \centering
  \includegraphics[width=\linewidth]{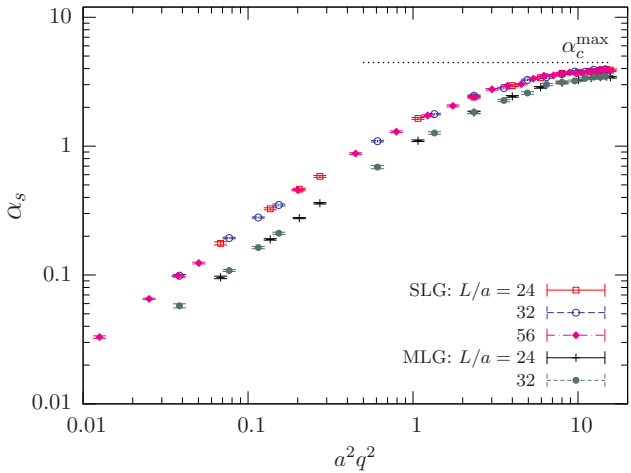}
  \caption{The running coupling, $\alpha_s$, for the standard (SLG)
    and modified (MLG) lattice Landau gauge at $\beta=0$. The dotted
    line, $\alpha^{\mathrm{max}}_c$, is the critical coupling in
    \Eq{eq:alphac} for $N_c=2$.}
  \label{fig:alpha_qq_beta0}
\end{figure}

In complete agreement with the general observation of conformal 
scaling at large momenta in the strong-coupling limit, the product
(\ref{alpha_minimom}) of the gluon and ghost dressing functions levels
at $\alpha_c \approx 4$ for large $a^2q^2$, as seen in
\Fig{fig:alpha_qq_beta0}. This is just below the upper bound
$\alpha^{\mathrm{max}}_c \approx 4.46$ for $SU(2)$.
The fact that $\alpha_c$ obtained here should be slightly smaller 
than this maximum value also complies with the continuum prediction
in Ref.~\cite{Lerche:2002ep}: Consistent with the results of the
previous section this is the expected trend for an exponent $\kappa$
which is slightly smaller than the  value in (\ref{kappa_c}).

It is furthermore quite compelling that this result, $\alpha_c $
close to $4$, is nearly independent of the
gauge-field definition, likewise. It is almost identical for SLG and
MLG, see \Fig{fig:alpha_qq_beta0}.  Predominantly driven by the ghost
propagator, the violations to conformal infrared scaling in the form
of a momentum dependence of $\alpha_s$ in the strong-coupling limit, 
 set in as soon as the ambiguity in the definition of minimal lattice
Landau gauge does.

%-----------------------------------------------------------------------
\subsection{Intermediate lattice couplings}
\label{sec:alphas_at_finBETA}

The significant differences observed at small momenta between SLG and
MLG so far were linked to the strong-coupling limit in which
discretization effects are enhanced to the extreme. Even though these
effects might be expected to disappear in the continuum limit,
eventually, it is important to assess to what extent they survive at
finite $\beta$. This is of relevance especially to the determinations
of the QCD scale parameter $\Lambda_{\overline{\mathsf{MS}}}\,$ based
on this running coupling in the perturbative domain from lattice
simulations \cite{Sternbeck:2007br,Boucaud:2008gn,Sternbeck:2010xu}.
In particular, it is of paramount importance to verify that this
ambiguity of minimal lattice Landau-gauge vanishes there, or else to
find ways to include it in the estimate of the systematic
uncertainties, if necessary.

\begin{figure}[t]
  \centering
  \includegraphics[width=\linewidth]{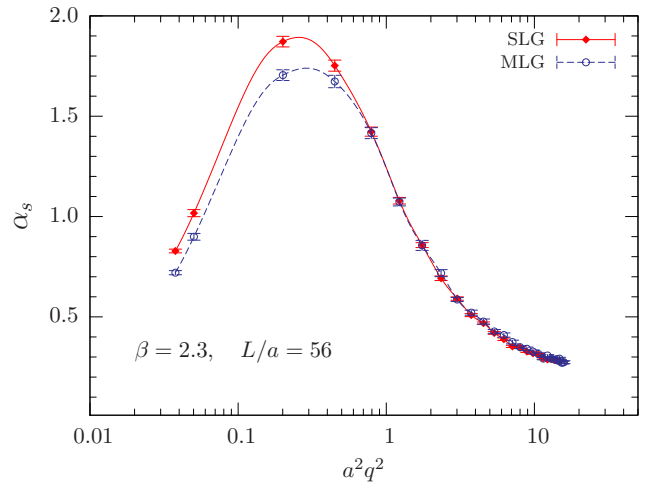}
  \caption{Data on $\alpha_s$ for the standard (SLG) and modified
    (MLG) lattice Landau gauge at $\beta=2.3$ using a $56^4$
    lattice. The lines are spline interpolations to guide the eye.}
  \label{fig:alpha_qq_beta2p3}
\end{figure}

As a first check, we have performed simulations in $SU(2)$ where gauge
configurations were generated with the one-plaquette Wilson action at
$\beta=2.3$ and~2.5. The configurations were then fixed, as above,
to SLG and to MLG, respectively, to measure and compare 
$\alpha_s$ on those two sets. For $\beta=2.3$ the results are shown in
\Fig{fig:alpha_qq_beta2p3}. Luckily, for the $\alpha_s$ project
mentioned above, we find no significant deviations between the two
definitions at large momenta. Both agree within the statistical errors
there, and the ambiguity does not appear to affect the high momentum
behavior of $\alpha_s(q^2) $ as relevant to the fits of its
perturbative expansion to extract the QCD scale. Of course this should
nevertheless be examined more carefully as one of the possible
systematic uncertainties also for the relevant $SU(3)$
configurations with dynamical quarks in future studies.

At lower momenta one observes significant differences between SLG and
MLG, especially in the transition region around the maxima at $a^2 q^2
\approx 0.25$. In units of the string-tension $\sigma$, with $a^2
\sigma = 0.145 $ for $\beta = 2.3$ from \cite{Langfeld:2007zw}, this
transition region corresponds to $q^2 \approx 1.7 \sigma $, or with
$\sigma = (440\, \mbox{MeV})^2$, in physical units, $ q \approx 570 $
MeV at maximum.  These observed differences persist for $\beta =
2.5$. A careful comparison of the strength of the effect for different
$\beta $ but comparable physical volumes, to disentangle
discretization and finite-volume effects at low momenta, is left for
future studies, however.

Because the Gribov copies of SLG and MLG differ, it seems quite
plausible that the observed ambiguity in the minimal lattice Landau
gauge beyond perturbation theory is closely related to the Gribov-copy
problem. This is also suggested by our data: the effect is
predominately due to the ghost propagator which is known to be
affected by this problem at low momenta
\cite{Cucchieri:1997dx,Bakeev:2003rr,Sternbeck:2005tk,Maas:2009ph}.

%--------------------------------------------------------------------------
\section{Summary and outlook}
\label{sec:conclusion}

We have studied gluon and ghost propagators of pure $SU(2)$ minimal
lattice Landau gauge theory in the strong-coupling limit. This
unphysical limit probes the gauge field measure of the minimal lattice
Landau gauge for there is no contribution from the Yang-Mills
(plaquette) action. The Faddeev-Popov determinant is implicitly
included by collapsing the gauge orbits onto the first Gribov region
as sampled by the minimal Landau gauge implementation used on the
lattice. The strong-coupling limit can therefore be thought of as a
means to implement, by hand, a lattice analogue of the infrared
dominance of ghost contributions in functional methods such as DSE or
FRGE studies.

As expected in the formal limit $ \Lambda_\mathrm{QCD} \to \infty$, it
is then observed that the dressing functions of both propagators, $Z$ and
$G$, show the conformal scaling behavior
\begin{displaymath}
  Z \propto (a^2q^2)^{2\kappa}
  \quad\text{and}\quad
  G \propto (a^2q^2)^{-\kappa}
\end{displaymath}
for large lattice momenta, $a^2 q^2 \gg 1$, well clear of the region
where finite-size effects should be expected. These effects, on the
other hand, turn out to be surprisingly small and the combined gluon and 
ghost data is consistent with an $L/a \to \infty $ extrapolation of a
critical exponent $\kappa = 0.57(3)$.  This scaling branch at large
$a^2q^2 $ furthermore leads to a critical coupling of $\alpha_c
\approx 4$ which is just below the predicted maximum
$\alpha^{\mathrm{max}}_c \approx 4.46$ for $SU(2)$. These results show
very little if no significant dependence on the lattice definition of
gauge fields and measure.

Another unambiguous result is the emergence of a transverse gluon mass
$M\propto 1/a$ in the strong-coupling limit of minimal lattice Landau
gauge. Both the gluon and ghost propagator show this massive behavior
at small momenta corresponding to 
\begin{displaymath} 
Z \sim q^2/M^2 \quad\text{and} \quad  G = \, \text{const.}  
\end{displaymath}
for $a^2q^2 \ll 1$. This massive low-momentum branch of the
data, however, depends strongly on which lattice definition is being
used for the gauge fields and their measure. This is typical for a
mass counter-term on the lattice and demonstrates the breakdown of
lattice Slavnov-Taylor identities (STIs) and BRST symmetry in minimal
lattice Landau gauge beyond perturbation theory.

It is still possible that this ambiguity disappears in the continuum
limit, eventually. But because it is a combination of ultraviolet
(mass counter-term) and infrared (breakdown of STIs) effects, this
might take very fine lattice spacings in combination with very large
volumes and therefore who-knows-how big lattices to verify explicitly.
As we have shown, the ambiguity is definitely present at commonly used
values of the lattice couplings in $SU(2)$.

It would obviously be desirable to have a BRST symmetry on the lattice
which could then provide lattice Slavnov-Taylor identities beyond
perturbation theory.  Non-perturbative lattice BRST has been plagued
by the Neuberger $0/0$ problem, but its improved topological
understanding provides ways to overcome this problem
\cite{vonSmekal:2008es}. It will be particularly interesting to see
whether the strong-coupling behavior of the propagators will change in
such approaches and whether this can lead to an unambiguous definition
of lattice Landau gauge beyond perturbation theory and in the
strong-coupling limit. Further studies and comparisons of different
approaches to non-perturbative gauge fixing on the lattice in two and
three dimensions, in particular in the strong-coupling limit will
thereby be valuable next steps. 

%--------------------------------------------------------------------------

\subsection*{Note added in revised version}

After completion of the original version of this paper, a follow-up
study of the strong-coupling limit
appeared~\cite{Cucchieri:2009zt}. There, independent    
data is presented for the SLG gluon and ghost propagators at $\beta=0$
in three and four dimensions. While the four dimensional data is in complete
agreement with the corresponding data presented here, the discussion
of our results in \cite{Cucchieri:2009zt} needs some clarification.

First, we have no intention to obscure the fact that all simulations
of gluon and ghost propagators using current lattice implementations
of the Landau gauge provide evidence for the qualitative low momentum
behavior of the decoupling solution (\ref{decoupling}) in three and
four dimensions. This includes all currently explored ranges for the
lattice size $L$ and coupling $\beta$, and our $\beta = 0$ results
are no exception.

In fact, in the strong-coupling limit the scaling solution would correspond to 
straight lines in double-logarithmic plots as indicated by the dashed/dotted
lines in Figs.~\ref{fig:gl_dress_qq_beta0-stdLG-U-Ud},
\ref{fig:gh_dress} and \ref{fig:alpha_qq_beta0}, possibly up to
finite-volume corrections which would gradually disappear with
increasing $L/a$. This is clearly not what we observe. Rather, the
plateaus in the data at small lattice momenta $ a^2q^2$ 
for both, the gluon propagator in
Fig.~\ref{fig:gl_dress_qq_beta0-stdLG-U-Ud} and the ghost dressing 
function in Fig.~\ref{fig:gh_dress}, 
are a clear indication of a massive (decoupling) behavior with almost
negligible finite-volume effects as we have demonstrated. 

It is nevertheless interesting, however, that we do observe for the first
time on the lattice a scaling behavior (\ref{eq:infrared-gh_gl})
with a scaling exponent  $\kappa$ right  inside
the expected range, in the large momentum regime of the
strong-coupling gluon propagator.
This has been confirmed by the data of Ref.~\cite{Cucchieri:2009zt}
and more recently also in \cite{Maas:2009ph}. 
The results for the ghost propagator are less conclusive but its
behavior at large lattice momenta tends to approach a form consistent
with the scaling relation (\ref{kappaZ=kappaG}), even though a clear
signature for scaling over a significant range of momenta is not observed.
The ghost propagator is most sensitive to the treatment of Gribov
copies and thus to the nonperturbative completion of the Landau
gauge. In fact, there is some recent evidence that the Gribov
ambiguity can be used to tune the ghost propagator to a scaling form
in the infrared \cite{Maas:2009se,Maas:2009ph}. This would identify
the one-parameter freedom in the continuum solutions to functional
equations \cite{Fischer:2008uz} as a second gauge parameter to
complete the Lorenz condition nonperturbatively.

It is an unambiguous result that without this tuning the
strong-coupling data for both propagators is well described by the
decoupling form at small momenta in lattice units, and by scaling at
large. To investigate the properties of the scaling branch at large
lattice momenta, we provide three different models to fit this branch
with results that are in good agreement of one another, and that give
a feeling for the systematic fit-model uncertainties at the same
time. In particular, the pure scaling model must fail when extending
the range of the fits too far beyond the scaling branch into the small
momentum region, and we see that explicitly happening in the corresponding
$\chi^2/\mathrm{ndf}$ in Fig.~\ref{fig:kappa_gl_lowbound}. An
alternative is to define effective exponents by logarithmic
derivatives and to monitor their momentum dependence
\cite{Cucchieri:2009zt}. This requires increased statistics but does
not change the general the properties of
the scaling branch and our conclusions \cite{Maas:2009ph}.
These does not imply, however, that we suggest to
dismiss the small momentum data, as we believe we were being
misinterpreted in Ref.~\cite{Cucchieri:2009zt}.  

This is in contrast to the authors of Ref.~\cite{Cucchieri:2009zt} who
write in their revised version that the data in the scaling region
should be discarded. They argue that the scale $a M \sim 1
$ for the transition between decoupling and scaling should be essentially 
given by $a\Lambda_\mathrm{QCD} $ and that momenta $p$ in the scaling
region therefore fail to satisfy $p \ll \Lambda_\mathrm{QCD}$. 
Even if the argument were correct, which it is not as we will explain,
we would not understand why any data should be discarded. It would
seem even more surprising and important to understand, if the
strong-coupling limit were to show infrared scaling for momenta on the
order of $\Lambda_\mathrm{QCD}$. Luckily that is not the case,
however. Here is the flaw in the argument: The authors of
\cite{Cucchieri:2009zt} use a recent lattice study of 
the phase diagram of QCD with one flavor of staggered quarks in the
strong-copuling limit \cite{deForcrand:2009dh} to assign `physical'
lengths to the lattice spacing which correspond to values of $a
\Lambda_\mathrm{QCD} $ between 0.62 and 1.92 depending on which hadron
mass or decay constant of the one-flavor strong-coupling model is
being used to set the absolute scale. There is no absolute scale in
the strong-coupling limit, however, not in the chiral limit of the
one-flavor model, let alone in the pure gauge theory. For the latter,
the relevant masses are those of the glueballs which all behave as $a
m \propto -\ln\beta \to \infty $ for $\beta \to 0$. It does make sense
to compare mass ratios, 
%for which the common prefactor $-\ln\beta$ cancels, 
but the relation between the absolute scale and $\Lambda_\mathrm{QCD}
$ of the scaling region is lost at $\beta = 0$. In units of the
latter, the only correct interpretation of the strong coupling limit
is $a\to \infty$. The fact that the strong-coupling mesons and baryons
are pointlike (in lattice units) actually demonstrates that also for the
quark model.
%\footnote{The range of the nuclear force is to be understood
%relative to the size of the nucleons at $\beta=0$ (which is $a$), likewise.}  

We moreover emphasize once more that the massive decoupling branch at small
lattice momenta is discretization dependent. We believe that this is
important and needs to be understood. It can hardly be due to
violations of rotational invariance as investigated, and not surprisingly
found to be very small, in \cite{Cucchieri:2009zt}. Note that we did
apply the usual cylinder cuts to our data in order to reduce those
effects as far as possible. Moreover, we only considered lattice
momenta with $a^2q^2\le 14$ in all our fits to the scaling branch as a
further precaution. To us the observed discretization dependences 
signal a breakdown of Slavnov-Taylor identities in minimal lattice
Landau gauge beyond perturbation theory as we explain. Maybe we are
over-interpreting the strong-coupling limit which unveiled this
problem, but the hope that it simply disappears in the
scaling region seems overly optimistic.  

\begin{figure}
  \centering
  \includegraphics[width=\linewidth]{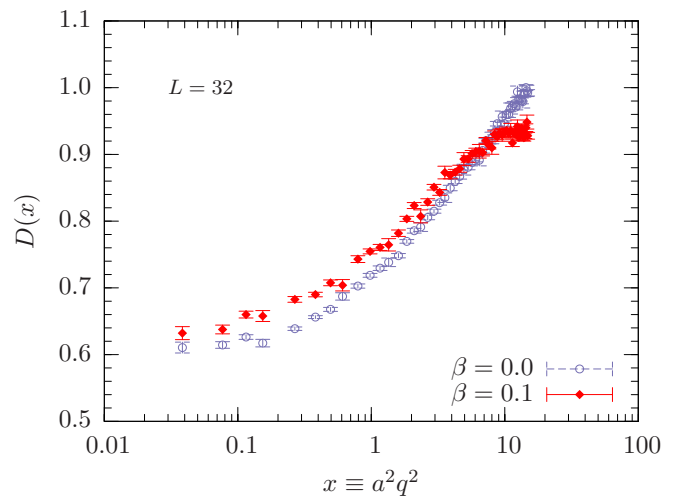}
  \caption{The gluon propagator in lattice units versus $a^2q^2$
    comparing data for $\beta=0$ and $\beta=0.1$. The $\beta=0$ data
    are the same as in \Fig{fig:gl_dress_qq_beta0-stdLG-U-Ud} ($32^4$
    lattice).}
  \label{fig:beta0vs0p1}
\end{figure}

The scaling properties such as exponent and coupling, on the other
hand, appear to be robust under variations of the discretization of
the gauge fields as far as we can judge from the available data.
It is also clear, however, that the scaling branch gets suppressed as soon as 
the gauge action is turned back on again, as demonstrated in
\Fig{fig:beta0vs0p1}, and that one is left with the decoupling branch
for low momenta at finite $\beta$. This emphasizes the importance of
understanding any discretization ambiguity of the associated gluon
mass, before concluding that this mass is now firmly
established.

%--------------------------------------------------------------------------
\section*{Acknowledgements}

This research was supported by the Australian Research Council and by
eResearch South Australia.\par

%---------------------------------------------------------------------------

\begin{appendix}

%--------------------------------------------------------------------------
\section{Standard lattice Landau gauge (SLG)}
\label{app:SLG}

In this Appendix we summarize the standard approach that is nowadays
widely adopted to study the gluon and ghost propagators of an SU(2)
[or SU(3)] Landau gauge theory using lattice simulations. Typically,
such simulations first generate a well-thermalized ensemble of
gauge-invariant link configurations which then, in a second step, is
fixed to what above we have called the standard lattice Landau gauge
(SLG). This is most conveniently done by employing either a
steepest-descent algorithm, like, e.g., the Fourier-accelerated
gauge-fixing of Refs.~\cite{Davies:1987vs,Cucchieri:1998ew}, or an
over-relaxation algorithm \cite{Mandula:1990vs}. Those iterative
algorithms are designed to relatively fast find (for a fixed $U$) a
set of local gauge transformations $U_{x\mu}\to
U^g_{x\mu}=g_xU_{x\mu}g^{\dagger}_{x+\hat{\mu}}$ that minimizes the
gauge functional of SLG which for the $SU(2)$ gauge group
is of the form
\begin{equation}
  V_U[g] = 4 \sum_{x,\mu}\left(1-\smallfrac{1}{2} \Tr
    U^g_{x\mu}\right)\;.
  \label{eq_app:functional}
\end{equation}
Any minimum of $V_U[g]$ automatically ensures that the (lattice)
Landau gauge condition, {\it i.e.}, the lattice (backward) derivative 
\begin{equation}
  F_x(A^g) \equiv \nabla^b_{\mu}A^g_{x\mu} = 0
  \label{eq_app:gaugecondition}
\end{equation}
is satisfied if the gauge-fixed (lattice) gluon fields are
given through the standard definition
\begin{equation}
  A_{x\mu}^g = \frac{1}{2ia}\left(U^g_{x\mu} -
    U^{g\,\dagger}_{x\mu}\right)
  \label{eq_app:gluonfield_SLG}
\end{equation}
in terms of the gauge-transformed link $U^g_{x\mu}$. Practically, in
order to fulfill  
condition~(\ref{eq_app:gaugecondition}) with sufficient accuracy, the
gauge-fixing algorithm is iterated until, e.g., the stopping criterion
\begin{equation}
  \label{eq_app:stoppingcriterion}
  \varepsilon =\max_x\;\Tr\big[(\nabla^b_{\mu} A_{x\mu}^g)(\nabla^b_{\mu}
  {A^{g\,\dagger}_{x\mu}}) \big]< 10^{-13}
\end{equation}
is satisfied for all lattice sites.

For the sake of completeness we mention that gauge functionals, like
\eqref{eq_app:functional}, do have a huge number of different local
minima. The corresponding gauge-fixed configurations do all satisfy
\Eq{eq_app:gaugecondition} and are related through local gauge
transformations to each other. This ambiguity, known as the Gribov-copy
problem on the lattice, causes minimization algorithms to find
different gauge-fixed configurations all with a different value
$V_U[g]$. This is, in particular, inevitable in the strong-coupling
limit. The ghost propagator is known to be affected by the Gribov-copy
problem while there seems to be no influence on the gluon propagator,
see, e.g., Refs.~\cite{Bakeev:2003rr,Sternbeck:2005tk} and
 for $\beta=0$ Ref.~\cite{Cucchieri:1997dx,Cucchieri:1997fy} in particular.

\medskip

To calculate the gluon propagator in momentum space, Fast Fourier
transformations (FFTs) are applied transforming the gauge-fixed gluon
fields $A_{x\mu}=A^a_{x\mu}\sigma^a/2$ (the superscript $g$
is dropped in what follows) into momentum space from which the gluon propagator
is then obtained for any momenta $k_{\mu}\in(N_{\mu}/2,N_{\mu}/2]$ as
the Monte-Carlo average
\begin{equation}
  \label{eq:gluon}
  D^{ab}_{\mu\nu}(k) = \left\langle A^a_{\mu}(k)A^b_{\nu}(-k) \right\rangle_U
\end{equation}
over gauge-fixed configurations~$U$. Since for the standard Wilson
gauge action the propagator's tree-level form on the lattice is of the
form
\begin{equation}
  \label{eq:treelevelgluon}
  D^{ab}_{0\,\mu\nu}(k) =
  \delta^{ab}\left(\delta_{\mu\nu}-\frac{q_{\mu}(k)q_{\nu}(k)}{q^2(k)}
  \right) \frac{1}{q^2(k)}
\end{equation}
where
\begin{equation}
  q_{\mu}(k) = \frac{2}{a}\sin\left(\frac{\pi k_{\mu}}{N_{\mu}}\right),
  \label{eq:physical_momenta}
\end{equation}
it is natural to associate $q_{\mu}$ with the physical momentum. Note
that then not only the continuum tensor structure of the gluon
propagator is retrieved but also the gauge condition is manifest in
momentum space, {\it i.e.}, 
\begin{equation}
  \sum_{\mu} q_{\mu}(k) A^a_{\mu}(k) = 0\;.
  \label{eq:gaugecondition_mom}
\end{equation}
The gluon dressing function is straightforwardly obtained from the
product $Z=q^2D(k)$ where $D(k)=D^{aa}_{\mu\mu}(k)/9$ (summing over
$a$ and $\mu$).

\medskip

Compared to the gluon propagator, determinations of the ghost
propagator 
\begin{displaymath}
  G^{ab}(k) = \sum_{xy}e^{ik(x-y)} \left\langle (M^{-1})^{ab}_{xy}\right\rangle_U
\end{displaymath}
are significantly more compute intensive as they involve inversions of
the Faddeev-Popov (F-P) operator $M$. On the lattice, $M$ is the
Hessian of the gauge functional $V_U[g]$ and for the $SU(2)$ SLG of the
form
\begin{equation}
  \label{eq:stdFP}
  \begin{split}
  M^{ab}_{xy}[U_{x\mu}] =  &\sum_{\mu}\bigg[\left(u^0_{x,\mu} +
    u^0_{x-\hat{\mu},\mu}\right)\delta^{ab}\delta_{xy}\\
  &-\,\left(u^0_{x,\mu}\delta^{ab} + \varepsilon^{abj}u^j_{x,\mu}\right)
     \delta_{x+\hat{\mu},y} \\
  &-\,\left(u^0_{x-\hat{\mu},\mu}\delta^{ab} -
  \varepsilon^{abj}u^j_{x-\hat{\mu},\mu}\right)\delta_{x-\hat{\mu},y}\bigg] 
  \end{split}
\end{equation}
where we have used the same notation as in \Eq{eq:quaternion_repr} and
assumed periodic boundary conditions. $M$ for the SU(3) SLG can be found,
e.g., in \cite{Sternbeck:2005tk}. 

Due to its zero eigenvalues, with corresponding constant eigenmodes,
it is impossible, however, to construct the inverse of
$M$. Nevertheless, all is not lost, as for the final analysis one is
usually interested only in data of $G$ at finite momenta for which $M$
can be inverted on a (non-constant) vector of plane waves
\hbox{$\xi^a_{x}(c,k)=\delta^{ac}e^{2\pi ikx}$} with
$k\neq0$. Typically, the momenta $k$ are chosen to survive both the
cylinder and cone cut \cite{Leinweber:1998uu} and then
$M^{ab}_{xy}\,\zeta^b_{y}(c,k) = \xi^a_{x}(c,k)$ is solved for
$\zeta(c,k)$ yielding the ghost propagator $G^{ab}(k)=\delta^{ab}G(k)$
(directly in momentum space) where $G(k)$ is the MC average
\begin{equation}
  \label{eq:ghost}
  G(k) = \frac{1}{3}
  \sum^3_{c=1}\left\langle\xi(k,c)\cdot\zeta(k,c)\right\rangle_U.
\end{equation}
In this way, also translational invariance is used to its full
capacity reducing the statistical noise to a minimum. 

For the inversion it is highly recommended to use a pre-conjugate
gradient algorithm, e.g., that of Ref.~\cite{Sternbeck:2005tk}. This
precondition has been proven to drastically accelerate computations,
in particular, when lattices sizes are as big as we have used for this
study and can also be straightforwardly applied to the case of Coulomb
gauge as done, e.g., in \cite{Voigt:2008rr,Nakagawa:2009zf}. For
further details on this technique refer to \cite{Sternbeck:2006rd}.

At tree-level the F-P operator is simply
$-\delta^{ab}\Delta_{xy}$ where $\Delta$ denotes the
lattice Laplacian in four dimensions. Consequently, the tree-level
form of the ghost propagator on the lattice is
\begin{equation}
  \label{eq:treelevelghost}
  G^{ab}_0(k) = - \delta^{ab}\frac{1}{q^2(k)}
\end{equation}
where $q_{\mu}$ is that of \Eq{eq:physical_momenta}.

%--------------------------------------------------------------------------
\section{Modified lattice Landau gauge (MLG)}
\label{app:MLG}

In this Appendix we summarize the Modified lattice Landau gauge (MLG)
of Ref.~\cite{vonSmekal:2007ns} and provide the reader with the
necessary information for the gauge group $SU(2)$ as used in
\Sec{sec:comparing_diff_latt_def}. 

The MLG is a novel implementation of the Landau gauge for lattice
gauge theories which was introduced in Ref.~\cite{vonSmekal:2007ns}
for the gauge groups $U(1)$ and $SU(2)$. It is based on stereographic  
projection to define lattice gauge fields and extensions to the gauge
group SU(3) or to Coulomb gauge are also possible.  

When comparing MLG to the ever popular SLG, there is no advantage that
the SLG has over the MLG. A promising particular feature of the MLG on
the other hand is that it provides a way to perform gauge-fixed MC 
simulations sampling \emph{all} Gribov copies of either sign (of the
Faddeev-Popov determinant) in the spirit of BRST. This is because in
MLG there is no perfect cancellation of Gribov copies of opposite
sign, known as the Neuberger 0/0 problem, which prevented us from
performing such simulations for 20 odd years. Note that in the
standard lattice approach to Landau gauge, {\it i.e.}, via a minimization
of the SLG functional [\Eq{eq_app:functional}], 
only Gribov copies within the first Gribov region, {\it i.e.}, with positive
sign, are sampled. Gauge-fixed Monte Carlo simulations via MLG,
however, would enable us to sample beyond that region. Indeed such
simulations are not meant to compete in any sense with standard MC
simulation of lattice gauge theory, as gauge-invariant observables
would be completely unaffected by that, but rather to provide a
theoretical sound framework for studying non-perturbative properties
of gauge-variant correlation functions.

This particular feature of MLG will be explored in a forthcoming
study. Here we simply have used the MLG for comparison in the standard
way, {\it i.e.}, we gauge-fix configurations via minimization of the
MLG functional,
\begin{equation}
  \label{eq_app:functional_modLG}
  \widetilde V_U[g] = -8 \sum_{x,\mu}
  \ln\left(\smallfrac{1}{2}+\smallfrac{1}{4}\Tr 
    U^{g}_{x\mu} \right). 
\end{equation}
Consequently, only the first Gribov region of MLG is sampled, however
when doing so we are in the fortunate position of having two
alternative lattice implementations of $SU(2)$ Landau gauge theory which
differ at finite lattice spacing, $a$, but meet in the continuum
limit. When comparing SLG to MLG data the impact of discretization
errors can then be seen at any finite $a$.

Leaving aside the issue of Gribov copies, we adapt the
Fourier-acceleration of Ref.~\cite{Davies:1987vs} to gauge-transform
configurations such that they satisfy the lattice Landau gauge
condition, here that of MLG,
\begin{equation}
  \nabla^b_{\mu}\widetilde{A}_{x\mu} = 0\;.
  \label{eq:gaugecondition_MLG}
\end{equation}
This automatically maximizes the MLG functional
[\Eq{eq_app:functional_modLG}] when $\widetilde{A}_{x\mu}$, the lattice
gluon field of MLG, is given through 
\begin{equation}
  \widetilde{A}_{x\mu} \equiv \frac{1}{2ia}\left(\widetilde{U}_{x\mu} -
    \widetilde{U}^{\dagger}_{x\mu}\right)
  \label{eq:gluonfield_MLG}
\end{equation}
where
\begin{equation}
  \widetilde{U}_{x\mu} 
     \equiv \frac{2U_{x\mu}}{1+\tfrac{1}{2} \Tr U_{x\mu}}\;.
 \label{eq:utilde}
\end{equation}
In our study the Fourier-accelerated gauge-fixing is iterated until the
stopping criterion
\begin{equation}
  \max_x\;\Tr\left[\nabla^b_{\mu} \widetilde{A}_{x\mu}\nabla^b_{\mu}
  \widetilde{A}^{\dagger}_{x\mu} \right] < 10^{-13}\;.
\label{eq_app:stopping_criterion_MLG}
\end{equation}
is met at all lattice sites.

Note that gauge-fixed configurations satisfying
\Eq{eq_app:stopping_criterion_MLG} do not satisfy that of SLG
[\Eq{eq_app:stoppingcriterion}] and vice versa. Nevertheless,
transversality of the corresponding lattice gluon field is ensured in
both cases, if momenta are associated with
\Eq{eq:physical_momenta} and the standard midpoint
definition is assumed. That is,
\begin{equation}
  \sum_{\mu} q_{\mu}(k) \widetilde{A}^a_{\mu}(k) = 0\;.
  \label{eq:gaugecondition_mom_MLG}
\end{equation}

The gluon propagator of MLG is straightforwardly constructed as in SLG
with $A$ substituted through $\widetilde{A}$. Similar holds for the ghost
propagator, though in MLG the F-P matrix is of the form
\begin{widetext}
\begin{align}
  \widetilde M^{ab}_{xy} = \sum_\mu \Bigg\{
  - &\Big( \tilde{u}_{x,\mu}^0
   \delta^{ab} + \epsilon^{abc} \tilde{u}_{x,\mu}^c 
   + \smallfrac{1}{2}\tilde{u}^a_{x,\mu}\tilde{u}^b_{x,\mu}\Big)
   \delta_{x+\hat{\mu},y}   
  +  \Big[( \tilde{u}_{x,\mu}^0 + \tilde{u}_{x-\hat{\mu},\mu}^0) \delta^{ab} +
  \smallfrac{1}{2}\tilde{u}^a_{x,\mu} \tilde{u}^b_{x,\mu}  \nonumber \\ 
  &+ \smallfrac{1}{2}\tilde{u}_{x-\hat{\mu},\mu}^a \tilde{u}_{x-\hat{\mu},\mu}^b
  \Big] \delta_{xy} - \Big( \tilde{u}_{x-\hat{\mu},\mu}^0 \delta^{ab} 
    - \epsilon^{abc} \tilde{u}_{x-\hat{\mu},\mu}^c +
   \smallfrac{1}{2}\tilde{u}^a_{x-\hat{\mu},\mu}\tilde{u}^b_{x-\hat{\mu},\mu}
   \Big) \delta_{x-\hat{\mu},y}\Bigg\}
\label{mod_M_FP}
\end{align}
\end{widetext}
where $\tilde{u}^0=2u^0/(1+u^0)$ and $\vec{u}=2\vec{u}/(1+u^0)$.
Due to the logarithm in \Eq{eq_app:functional_modLG} there are $\tilde{u}^a
\tilde{u}^b$-terms quadratic in the projected variables
$\tilde{u}$. Apart from those, $\widetilde{M}$ is of the same form as
the standard F-P operator in $SU(2)$ [\Eq{eq:stdFP}] with the $u$'s replaced
by $\tilde{u}$. Therefore, the tree-level forms of the gluon and
ghost propagators, \Eq{eq:treelevelgluon} and \eqref{eq:treelevelghost}, are
also valid in MLG.

\end{appendix}

%Added the SLACcitation info to the apsrev.bst style (for details consult
% http://www-spires.fnal.gov/spires/hep/refs/bibstyles.shtml)..Andre
\bibliographystyle{apsrev-SLAC}
\bibliography{references}

\begin{thebibliography}{63}
\expandafter\ifx\csname natexlab\endcsname\relax\def\natexlab#1{#1}\fi
\expandafter\ifx\csname bibnamefont\endcsname\relax
  \def\bibnamefont#1{#1}\fi
\expandafter\ifx\csname bibfnamefont\endcsname\relax
  \def\bibfnamefont#1{#1}\fi
\expandafter\ifx\csname citenamefont\endcsname\relax
  \def\citenamefont#1{#1}\fi
\expandafter\ifx\csname url\endcsname\relax
  \def\url#1{\texttt{#1}}\fi
\expandafter\ifx\csname urlprefix\endcsname\relax\def\urlprefix{URL }\fi
\providecommand{\bibinfo}[2]{#2}
\providecommand{\eprint}[2][]{\url{#2}}

\bibitem[{\citenamefont{Roberts and Williams}(1994)}]{Roberts:1994dr}
\bibinfo{author}{\bibfnamefont{C.~D.} \bibnamefont{Roberts}} \bibnamefont{and}
  \bibinfo{author}{\bibfnamefont{A.~G.} \bibnamefont{Williams}},
  \bibinfo{journal}{Prog. Part. Nucl. Phys.} \textbf{\bibinfo{volume}{33}},
  \bibinfo{pages}{477} (\bibinfo{year}{1994}), \eprint{hep-ph/9403224}.
%%CITATION = HEP-PH/9403224;%%

\bibitem[{\citenamefont{Alkofer and von Smekal}(2001)}]{Alkofer:2000wg}
\bibinfo{author}{\bibfnamefont{R.}~\bibnamefont{Alkofer}} \bibnamefont{and}
  \bibinfo{author}{\bibfnamefont{L.}~\bibnamefont{von Smekal}},
  \bibinfo{journal}{Phys. Rept.} \textbf{\bibinfo{volume}{353}},
  \bibinfo{pages}{281} (\bibinfo{year}{2001}), \eprint{hep-ph/0007355}.
%%CITATION = HEP-PH/0007355;%%

\bibitem[{\citenamefont{Fischer}(2006)}]{Fischer:2006ub}
\bibinfo{author}{\bibfnamefont{C.~S.} \bibnamefont{Fischer}},
  \bibinfo{journal}{J. Phys.} \textbf{\bibinfo{volume}{G32}},
  \bibinfo{pages}{R253} (\bibinfo{year}{2006}), \eprint{hep-ph/0605173}.
%%CITATION = HEP-PH/0605173;%%

\bibitem[{\citenamefont{Roberts et~al.}(2007)\citenamefont{Roberts, Bhagwat,
  Holl, and Wright}}]{Roberts:2007jh}
\bibinfo{author}{\bibfnamefont{C.~D.} \bibnamefont{Roberts}},
  \bibinfo{author}{\bibfnamefont{M.~S.} \bibnamefont{Bhagwat}},
  \bibinfo{author}{\bibfnamefont{A.}~\bibnamefont{Holl}}, \bibnamefont{and}
  \bibinfo{author}{\bibfnamefont{S.~V.} \bibnamefont{Wright}},
  \bibinfo{journal}{Eur. Phys. J. ST} \textbf{\bibinfo{volume}{140}},
  \bibinfo{pages}{53} (\bibinfo{year}{2007}), \eprint{0802.0217}.
%%CITATION = 0802.0217;%%

\bibitem[{\citenamefont{von Smekal et~al.}(1997)\citenamefont{von Smekal,
  Alkofer, and Hauck}}]{vonSmekal:1997is}
\bibinfo{author}{\bibfnamefont{L.}~\bibnamefont{von Smekal}},
  \bibinfo{author}{\bibfnamefont{R.}~\bibnamefont{Alkofer}}, \bibnamefont{and}
  \bibinfo{author}{\bibfnamefont{A.}~\bibnamefont{Hauck}},
  \bibinfo{journal}{Phys. Rev. Lett.} \textbf{\bibinfo{volume}{79}},
  \bibinfo{pages}{3591} (\bibinfo{year}{1997}), \eprint{hep-ph/9705242}.
%%CITATION = HEP-PH 9705242;%%

\bibitem[{\citenamefont{von Smekal et~al.}(1998)\citenamefont{von Smekal,
  Hauck, and Alkofer}}]{vonSmekal:1997vx}
\bibinfo{author}{\bibfnamefont{L.}~\bibnamefont{von Smekal}},
  \bibinfo{author}{\bibfnamefont{A.}~\bibnamefont{Hauck}}, \bibnamefont{and}
  \bibinfo{author}{\bibfnamefont{R.}~\bibnamefont{Alkofer}},
  \bibinfo{journal}{Ann. Phys.} \textbf{\bibinfo{volume}{267}},
  \bibinfo{pages}{1} (\bibinfo{year}{1998}), \eprint{hep-ph/9707327}.
%%CITATION = HEP-PH 9707327;%%

\bibitem[{\citenamefont{Alkofer and von Smekal}(2000)}]{Alkofer:2000mz}
\bibinfo{author}{\bibfnamefont{R.}~\bibnamefont{Alkofer}} \bibnamefont{and}
  \bibinfo{author}{\bibfnamefont{L.}~\bibnamefont{von Smekal}},
  \bibinfo{journal}{Nucl. Phys.} \textbf{\bibinfo{volume}{A680}},
  \bibinfo{pages}{133} (\bibinfo{year}{2000}), \eprint{hep-ph/0004141}.
%%CITATION = HEP-PH/0004141;%%

\bibitem[{\citenamefont{Lerche and von Smekal}(2002)}]{Lerche:2002ep}
\bibinfo{author}{\bibfnamefont{C.}~\bibnamefont{Lerche}} \bibnamefont{and}
  \bibinfo{author}{\bibfnamefont{L.}~\bibnamefont{von Smekal}},
  \bibinfo{journal}{Phys. Rev.} \textbf{\bibinfo{volume}{D65}},
  \bibinfo{pages}{125006} (\bibinfo{year}{2002}), \eprint{hep-ph/0202194}.
%%CITATION = HEP-PH/0202194;%%

\bibitem[{\citenamefont{Zwanziger}(2002)}]{Zwanziger:2001kw}
\bibinfo{author}{\bibfnamefont{D.}~\bibnamefont{Zwanziger}},
  \bibinfo{journal}{Phys. Rev.} \textbf{\bibinfo{volume}{D65}},
  \bibinfo{pages}{094039} (\bibinfo{year}{2002}), \eprint{hep-th/0109224}.
%%CITATION = HEP-TH/0109224;%%

\bibitem[{\citenamefont{Pawlowski et~al.}(2004)\citenamefont{Pawlowski, Litim,
  Nedelko, and von Smekal}}]{Pawlowski:2003hq}
\bibinfo{author}{\bibfnamefont{J.~M.} \bibnamefont{Pawlowski}},
  \bibinfo{author}{\bibfnamefont{D.~F.} \bibnamefont{Litim}},
  \bibinfo{author}{\bibfnamefont{S.}~\bibnamefont{Nedelko}}, \bibnamefont{and}
  \bibinfo{author}{\bibfnamefont{L.}~\bibnamefont{von Smekal}},
  \bibinfo{journal}{Phys. Rev. Lett.} \textbf{\bibinfo{volume}{93}},
  \bibinfo{pages}{152002} (\bibinfo{year}{2004}), \eprint{hep-th/0312324}.
%%CITATION = HEP-TH/0312324;%%

\bibitem[{\citenamefont{Maas}(2007)}]{Maas:2007uv}
\bibinfo{author}{\bibfnamefont{A.}~\bibnamefont{Maas}}, \bibinfo{journal}{Phys.
  Rev.} \textbf{\bibinfo{volume}{D75}}, \bibinfo{pages}{116004}
  (\bibinfo{year}{2007}), \eprint{0704.0722}.
%%CITATION = 0704.0722;%%

\bibitem[{\citenamefont{Sternbeck et~al.}(2006)\citenamefont{Sternbeck,
  Ilgenfritz, M{\"u}ller-Preussker, Schiller, and
  Bogolubsky}}]{Sternbeck:2006cg}
\bibinfo{author}{\bibfnamefont{A.}~\bibnamefont{Sternbeck}},
  \bibinfo{author}{\bibfnamefont{E.-M.} \bibnamefont{Ilgenfritz}},
  \bibinfo{author}{\bibfnamefont{M.}~\bibnamefont{M{\"u}ller-Preussker}},
  \bibinfo{author}{\bibfnamefont{A.}~\bibnamefont{Schiller}}, \bibnamefont{and}
  \bibinfo{author}{\bibfnamefont{I.~L.} \bibnamefont{Bogolubsky}},
  \bibinfo{journal}{PoS} \textbf{\bibinfo{volume}{LAT2006}},
  \bibinfo{pages}{076} (\bibinfo{year}{2006}), \eprint{hep-lat/0610053}.
%%CITATION = HEP-LAT/0610053;%%

\bibitem[{\citenamefont{Ilgenfritz et~al.}(2007)\citenamefont{Ilgenfritz,
  M{\"u}ller-Preussker, Sternbeck, Schiller, and
  Bogolubsky}}]{Ilgenfritz:2006he}
\bibinfo{author}{\bibfnamefont{E.-M.} \bibnamefont{Ilgenfritz}},
  \bibinfo{author}{\bibfnamefont{M.}~\bibnamefont{M{\"u}ller-Preussker}},
  \bibinfo{author}{\bibfnamefont{A.}~\bibnamefont{Sternbeck}},
  \bibinfo{author}{\bibfnamefont{A.}~\bibnamefont{Schiller}}, \bibnamefont{and}
  \bibinfo{author}{\bibfnamefont{I.~L.} \bibnamefont{Bogolubsky}},
  \bibinfo{journal}{Braz. J. Phys.} \textbf{\bibinfo{volume}{37}},
  \bibinfo{pages}{193} (\bibinfo{year}{2007}), \eprint{hep-lat/0609043}.
%%CITATION = HEP-LAT/0609043;%%

\bibitem[{\citenamefont{Sternbeck
  et~al.}(2007{\natexlab{a}})\citenamefont{Sternbeck, von Smekal, Leinweber,
  and Williams}}]{Sternbeck:2007ug}
\bibinfo{author}{\bibfnamefont{A.}~\bibnamefont{Sternbeck}},
  \bibinfo{author}{\bibfnamefont{L.}~\bibnamefont{von Smekal}},
  \bibinfo{author}{\bibfnamefont{D.~B.} \bibnamefont{Leinweber}},
  \bibnamefont{and} \bibinfo{author}{\bibfnamefont{A.~G.}
  \bibnamefont{Williams}}, \bibinfo{journal}{PoS}
  \textbf{\bibinfo{volume}{LAT2007}}, \bibinfo{pages}{340}
  (\bibinfo{year}{2007}{\natexlab{a}}), \eprint{0710.1982}.
%%CITATION = 0710.1982;%%

\bibitem[{\citenamefont{Cucchieri and Mendes}(2007)}]{Cucchieri:2007md}
\bibinfo{author}{\bibfnamefont{A.}~\bibnamefont{Cucchieri}} \bibnamefont{and}
  \bibinfo{author}{\bibfnamefont{T.}~\bibnamefont{Mendes}},
  \bibinfo{journal}{PoS} \textbf{\bibinfo{volume}{LAT2007}},
  \bibinfo{pages}{297} (\bibinfo{year}{2007}), \eprint{0710.0412}.
%%CITATION = 0710.0412;%%

\bibitem[{\citenamefont{Bogolubsky et~al.}(2007)\citenamefont{Bogolubsky,
  Ilgenfritz, M{\"u}ller-Preussker, and Sternbeck}}]{Bogolubsky:2007ud}
\bibinfo{author}{\bibfnamefont{I.~L.} \bibnamefont{Bogolubsky}},
  \bibinfo{author}{\bibfnamefont{E.-M.} \bibnamefont{Ilgenfritz}},
  \bibinfo{author}{\bibfnamefont{M.}~\bibnamefont{M{\"u}ller-Preussker}},
  \bibnamefont{and}
  \bibinfo{author}{\bibfnamefont{A.}~\bibnamefont{Sternbeck}},
  \bibinfo{journal}{PoS} \textbf{\bibinfo{volume}{LAT2007}},
  \bibinfo{pages}{290} (\bibinfo{year}{2007}), \eprint{0710.1968}.
%%CITATION = 0710.1968;%%

\bibitem[{\citenamefont{Cucchieri and
  Mendes}(2008{\natexlab{a}})}]{Cucchieri:2007rg}
\bibinfo{author}{\bibfnamefont{A.}~\bibnamefont{Cucchieri}} \bibnamefont{and}
  \bibinfo{author}{\bibfnamefont{T.}~\bibnamefont{Mendes}},
  \bibinfo{journal}{Phys. Rev. Lett.} \textbf{\bibinfo{volume}{100}},
  \bibinfo{pages}{241601} (\bibinfo{year}{2008}{\natexlab{a}}),
  \eprint{0712.3517}.
%%CITATION = 0712.3517;%%

\bibitem[{\citenamefont{Cucchieri and
  Mendes}(2008{\natexlab{b}})}]{Cucchieri:2008fc}
\bibinfo{author}{\bibfnamefont{A.}~\bibnamefont{Cucchieri}} \bibnamefont{and}
  \bibinfo{author}{\bibfnamefont{T.}~\bibnamefont{Mendes}},
  \bibinfo{journal}{Phys. Rev.} \textbf{\bibinfo{volume}{D78}},
  \bibinfo{pages}{094503} (\bibinfo{year}{2008}{\natexlab{b}}),
  \eprint{0804.2371}.
%%CITATION = 0804.2371;%%

\bibitem[{\citenamefont{Bogolubsky et~al.}(2009)\citenamefont{Bogolubsky,
  Ilgenfritz, M{\"u}ller-Preussker, and Sternbeck}}]{Bogolubsky:2009dc}
\bibinfo{author}{\bibfnamefont{I.~L.} \bibnamefont{Bogolubsky}},
  \bibinfo{author}{\bibfnamefont{E.-M.} \bibnamefont{Ilgenfritz}},
  \bibinfo{author}{\bibfnamefont{M.}~\bibnamefont{M{\"u}ller-Preussker}},
  \bibnamefont{and}
  \bibinfo{author}{\bibfnamefont{A.}~\bibnamefont{Sternbeck}},
  \bibinfo{journal}{Phys. Lett.} \textbf{\bibinfo{volume}{B676}},
  \bibinfo{pages}{69} (\bibinfo{year}{2009}), \eprint{0901.0736}.
%%CITATION = 0901.0736;%%

\bibitem[{\citenamefont{Aguilar and Natale}(2004)}]{Aguilar:2004sw}
\bibinfo{author}{\bibfnamefont{A.~C.} \bibnamefont{Aguilar}} \bibnamefont{and}
  \bibinfo{author}{\bibfnamefont{A.~A.} \bibnamefont{Natale}},
  \bibinfo{journal}{JHEP} \textbf{\bibinfo{volume}{08}}, \bibinfo{pages}{057}
  (\bibinfo{year}{2004}), \eprint{hep-ph/0408254}.
%%CITATION = HEP-PH/0408254;%%

\bibitem[{\citenamefont{Boucaud et~al.}(2006)}]{Boucaud:2006if}
\bibinfo{author}{\bibfnamefont{P.}~\bibnamefont{Boucaud}} \bibnamefont{et~al.},
  \bibinfo{journal}{JHEP} \textbf{\bibinfo{volume}{06}}, \bibinfo{pages}{001}
  (\bibinfo{year}{2006}), \eprint{hep-ph/0604056}.
%%CITATION = HEP-PH/0604056;%%

\bibitem[{\citenamefont{Dudal et~al.}(2008{\natexlab{a}})\citenamefont{Dudal,
  Sorella, Vandersickel, and Verschelde}}]{Dudal:2007cw}
\bibinfo{author}{\bibfnamefont{D.}~\bibnamefont{Dudal}},
  \bibinfo{author}{\bibfnamefont{S.~P.} \bibnamefont{Sorella}},
  \bibinfo{author}{\bibfnamefont{N.}~\bibnamefont{Vandersickel}},
  \bibnamefont{and}
  \bibinfo{author}{\bibfnamefont{H.}~\bibnamefont{Verschelde}},
  \bibinfo{journal}{Phys. Rev.} \textbf{\bibinfo{volume}{D77}},
  \bibinfo{pages}{071501} (\bibinfo{year}{2008}{\natexlab{a}}),
  \eprint{0711.4496}.
%%CITATION = 0711.4496;%%

\bibitem[{\citenamefont{Aguilar
  et~al.}(2008{\natexlab{a}})\citenamefont{Aguilar, Binosi, and
  Papavassiliou}}]{Aguilar:2008xm}
\bibinfo{author}{\bibfnamefont{A.~C.} \bibnamefont{Aguilar}},
  \bibinfo{author}{\bibfnamefont{D.}~\bibnamefont{Binosi}}, \bibnamefont{and}
  \bibinfo{author}{\bibfnamefont{J.}~\bibnamefont{Papavassiliou}},
  \bibinfo{journal}{Phys. Rev.} \textbf{\bibinfo{volume}{D78}},
  \bibinfo{pages}{025010} (\bibinfo{year}{2008}{\natexlab{a}}),
  \eprint{0802.1870}.
%%CITATION = 0802.1870;%%

\bibitem[{\citenamefont{Aguilar
  et~al.}(2008{\natexlab{b}})\citenamefont{Aguilar, Binosi, and
  Papavassiliou}}]{Aguilar:2008fh}
\bibinfo{author}{\bibfnamefont{A.~C.} \bibnamefont{Aguilar}},
  \bibinfo{author}{\bibfnamefont{D.}~\bibnamefont{Binosi}}, \bibnamefont{and}
  \bibinfo{author}{\bibfnamefont{J.}~\bibnamefont{Papavassiliou}},
  \bibinfo{journal}{PoS} \textbf{\bibinfo{volume}{LC2008}},
  \bibinfo{pages}{050} (\bibinfo{year}{2008}{\natexlab{b}}),
  \eprint{0810.2333}.
%%CITATION = 0810.2333;%%

\bibitem[{\citenamefont{Boucaud et~al.}(2008{\natexlab{a}})}]{Boucaud:2008ji}
\bibinfo{author}{\bibfnamefont{P.}~\bibnamefont{Boucaud}} \bibnamefont{et~al.},
  \bibinfo{journal}{JHEP} \textbf{\bibinfo{volume}{06}}, \bibinfo{pages}{012}
  (\bibinfo{year}{2008}{\natexlab{a}}), \eprint{0801.2721}.
%%CITATION = 0801.2721;%%

\bibitem[{\citenamefont{Boucaud et~al.}(2008{\natexlab{b}})}]{Boucaud:2008ky}
\bibinfo{author}{\bibfnamefont{P.}~\bibnamefont{Boucaud}} \bibnamefont{et~al.},
  \bibinfo{journal}{JHEP} \textbf{\bibinfo{volume}{06}}, \bibinfo{pages}{099}
  (\bibinfo{year}{2008}{\natexlab{b}}), \eprint{0803.2161}.
%%CITATION = 0803.2161;%%

\bibitem[{\citenamefont{Fischer et~al.}(2009)\citenamefont{Fischer, Maas, and
  Pawlowski}}]{Fischer:2008uz}
\bibinfo{author}{\bibfnamefont{C.~S.} \bibnamefont{Fischer}},
  \bibinfo{author}{\bibfnamefont{A.}~\bibnamefont{Maas}}, \bibnamefont{and}
  \bibinfo{author}{\bibfnamefont{J.~M.} \bibnamefont{Pawlowski}},
  \bibinfo{journal}{Annals Phys.} \textbf{\bibinfo{volume}{324}},
  \bibinfo{pages}{2408} (\bibinfo{year}{2009}), \eprint{0810.1987}.
%%CITATION = 0810.1987;%%

\bibitem[{\citenamefont{Sternbeck and von
  Smekal}(2008{\natexlab{a}})}]{Sternbeck:2008wg}
\bibinfo{author}{\bibfnamefont{A.}~\bibnamefont{Sternbeck}} \bibnamefont{and}
  \bibinfo{author}{\bibfnamefont{L.}~\bibnamefont{von Smekal}},
  \bibinfo{journal}{PoS} \textbf{\bibinfo{volume}{LATTICE2008}},
  \bibinfo{pages}{267} (\bibinfo{year}{2008}{\natexlab{a}}),
  \eprint{0810.3765}.
%%CITATION = 0810.3765;%%

\bibitem[{\citenamefont{Sternbeck and von
  Smekal}(2008{\natexlab{b}})}]{Sternbeck:2008na}
\bibinfo{author}{\bibfnamefont{A.}~\bibnamefont{Sternbeck}} \bibnamefont{and}
  \bibinfo{author}{\bibfnamefont{L.}~\bibnamefont{von Smekal}},
  \bibinfo{journal}{PoS} \textbf{\bibinfo{volume}{CONFINEMENT8}},
  \bibinfo{pages}{049} (\bibinfo{year}{2008}{\natexlab{b}}),
  \eprint{0812.3268}.
%%CITATION = 0812.3268;%%

\bibitem[{\citenamefont{Gribov}(1978)}]{Gribov:1977wm}
\bibinfo{author}{\bibfnamefont{V.~N.} \bibnamefont{Gribov}},
  \bibinfo{journal}{Nucl. Phys.} \textbf{\bibinfo{volume}{B139}},
  \bibinfo{pages}{1} (\bibinfo{year}{1978}).
%%CITATION = NUPHA,B139,1;%%

\bibitem[{\citenamefont{Alkofer et~al.}(2005)\citenamefont{Alkofer, Fischer,
  and Llanes-Estrada}}]{Alkofer:2004it}
\bibinfo{author}{\bibfnamefont{R.}~\bibnamefont{Alkofer}},
  \bibinfo{author}{\bibfnamefont{C.~S.} \bibnamefont{Fischer}},
  \bibnamefont{and} \bibinfo{author}{\bibfnamefont{F.~J.}
  \bibnamefont{Llanes-Estrada}}, \bibinfo{journal}{Phys. Lett.}
  \textbf{\bibinfo{volume}{B611}}, \bibinfo{pages}{279} (\bibinfo{year}{2005}),
  \eprint{hep-th/0412330}.
%%CITATION = HEP-TH/0412330;%%

\bibitem[{\citenamefont{Taylor}(1971)}]{Taylor:1971ff}
\bibinfo{author}{\bibfnamefont{J.~C.} \bibnamefont{Taylor}},
  \bibinfo{journal}{Nucl. Phys.} \textbf{\bibinfo{volume}{B33}},
  \bibinfo{pages}{436} (\bibinfo{year}{1971}).
%%CITATION = NUPHA,B33,436;%%

\bibitem[{\citenamefont{Fischer and Pawlowski}(2007)}]{Fischer:2006vf}
\bibinfo{author}{\bibfnamefont{C.~S.} \bibnamefont{Fischer}} \bibnamefont{and}
  \bibinfo{author}{\bibfnamefont{J.~M.} \bibnamefont{Pawlowski}},
  \bibinfo{journal}{Phys. Rev.} \textbf{\bibinfo{volume}{D75}},
  \bibinfo{pages}{025012} (\bibinfo{year}{2007}), \eprint{hep-th/0609009}.
%%CITATION = HEP-TH/0609009;%%

\bibitem[{\citenamefont{Fischer and Pawlowski}(2009)}]{Fischer:2009tn}
\bibinfo{author}{\bibfnamefont{C.~S.} \bibnamefont{Fischer}} \bibnamefont{and}
  \bibinfo{author}{\bibfnamefont{J.~M.} \bibnamefont{Pawlowski}},
  \bibinfo{journal}{Phys. Rev.} \textbf{\bibinfo{volume}{D80}},
  \bibinfo{pages}{025023} (\bibinfo{year}{2009}), \eprint{0903.2193}.
%%CITATION = 0903.2193;%%

\bibitem[{\citenamefont{von Smekal et~al.}(2009)\citenamefont{von Smekal,
  Maltman, and Sternbeck}}]{vonSmekal:2009ae}
\bibinfo{author}{\bibfnamefont{L.}~\bibnamefont{von Smekal}},
  \bibinfo{author}{\bibfnamefont{K.}~\bibnamefont{Maltman}}, \bibnamefont{and}
  \bibinfo{author}{\bibfnamefont{A.}~\bibnamefont{Sternbeck}},
  \bibinfo{journal}{Phys. Lett.} \textbf{\bibinfo{volume}{B681}},
  \bibinfo{pages}{336} (\bibinfo{year}{2009}), \eprint{0903.1696}.
%%CITATION = 0903.1696;%%

\bibitem[{\citenamefont{Fischer et~al.}(2007)\citenamefont{Fischer, Maas,
  Pawlowski, and von Smekal}}]{Fischer:2007pf}
\bibinfo{author}{\bibfnamefont{C.~S.} \bibnamefont{Fischer}},
  \bibinfo{author}{\bibfnamefont{A.}~\bibnamefont{Maas}},
  \bibinfo{author}{\bibfnamefont{J.~M.} \bibnamefont{Pawlowski}},
  \bibnamefont{and} \bibinfo{author}{\bibfnamefont{L.}~\bibnamefont{von
  Smekal}}, \bibinfo{journal}{Annals Phys.} \textbf{\bibinfo{volume}{322}},
  \bibinfo{pages}{2916} (\bibinfo{year}{2007}), \eprint{hep-ph/0701050}.
%%CITATION = HEP-PH/0701050;%%

\bibitem[{\citenamefont{von Smekal}(2008)}]{vonSmekal:2008ws}
\bibinfo{author}{\bibfnamefont{L.}~\bibnamefont{von Smekal}}
  (\bibinfo{year}{2008}), \eprint{0812.0654}.
%%CITATION = 0812.0654;%%

\bibitem[{\citenamefont{Zwanziger}(1993)}]{Zwanziger:1992qr}
\bibinfo{author}{\bibfnamefont{D.}~\bibnamefont{Zwanziger}},
  \bibinfo{journal}{Nucl. Phys.} \textbf{\bibinfo{volume}{B399}},
  \bibinfo{pages}{477} (\bibinfo{year}{1993}).
%%CITATION = NUPHA,B399,477;%%

\bibitem[{\citenamefont{Dudal et~al.}(2008{\natexlab{b}})\citenamefont{Dudal,
  Gracey, Sorella, Vandersickel, and Verschelde}}]{Dudal:2008sp}
\bibinfo{author}{\bibfnamefont{D.}~\bibnamefont{Dudal}},
  \bibinfo{author}{\bibfnamefont{J.~A.} \bibnamefont{Gracey}},
  \bibinfo{author}{\bibfnamefont{S.~P.} \bibnamefont{Sorella}},
  \bibinfo{author}{\bibfnamefont{N.}~\bibnamefont{Vandersickel}},
  \bibnamefont{and}
  \bibinfo{author}{\bibfnamefont{H.}~\bibnamefont{Verschelde}},
  \bibinfo{journal}{Phys. Rev.} \textbf{\bibinfo{volume}{D78}},
  \bibinfo{pages}{065047} (\bibinfo{year}{2008}{\natexlab{b}}),
  \eprint{0806.4348}.
%%CITATION = 0806.4348;%%

\bibitem[{\citenamefont{von Smekal et~al.}(2008{\natexlab{a}})\citenamefont{von
  Smekal, Ghiotti, and Williams}}]{vonSmekal:2008en}
\bibinfo{author}{\bibfnamefont{L.}~\bibnamefont{von Smekal}},
  \bibinfo{author}{\bibfnamefont{M.}~\bibnamefont{Ghiotti}}, \bibnamefont{and}
  \bibinfo{author}{\bibfnamefont{A.~G.} \bibnamefont{Williams}},
  \bibinfo{journal}{Phys. Rev.} \textbf{\bibinfo{volume}{D78}},
  \bibinfo{pages}{085016} (\bibinfo{year}{2008}{\natexlab{a}}),
  \eprint{0807.0480}.
%%CITATION = 0807.0480;%%

\bibitem[{\citenamefont{Braun et~al.}(2010)\citenamefont{Braun, Gies, and
  Pawlowski}}]{Braun:2007bx}
\bibinfo{author}{\bibfnamefont{J.}~\bibnamefont{Braun}},
  \bibinfo{author}{\bibfnamefont{H.}~\bibnamefont{Gies}}, \bibnamefont{and}
  \bibinfo{author}{\bibfnamefont{J.~M.} \bibnamefont{Pawlowski}},
  \bibinfo{journal}{Phys. Lett.} \textbf{\bibinfo{volume}{B684}},
  \bibinfo{pages}{262} (\bibinfo{year}{2010}), \eprint{0708.2413}.
%%CITATION = 0708.2413;%%

\bibitem[{\citenamefont{Maas et~al.}(2010)\citenamefont{Maas, Pawlowski,
  Spielmann, Sternbeck, and von Smekal}}]{Maas:2009ph}
\bibinfo{author}{\bibfnamefont{A.}~\bibnamefont{Maas}},
  \bibinfo{author}{\bibfnamefont{J.~M.} \bibnamefont{Pawlowski}},
  \bibinfo{author}{\bibfnamefont{D.}~\bibnamefont{Spielmann}},
  \bibinfo{author}{\bibfnamefont{A.}~\bibnamefont{Sternbeck}},
  \bibnamefont{and} \bibinfo{author}{\bibfnamefont{L.}~\bibnamefont{von
  Smekal}}, \bibinfo{journal}{Eur. Phys. J. C, in press}
  (\bibinfo{year}{2010}), \eprint{0912.4203}.
%%CITATION = 0912.4203;%%

\bibitem[{\citenamefont{von Smekal et~al.}(2007)\citenamefont{von Smekal,
  Mehta, Sternbeck, and Williams}}]{vonSmekal:2007ns}
\bibinfo{author}{\bibfnamefont{L.}~\bibnamefont{von Smekal}},
  \bibinfo{author}{\bibfnamefont{D.}~\bibnamefont{Mehta}},
  \bibinfo{author}{\bibfnamefont{A.}~\bibnamefont{Sternbeck}},
  \bibnamefont{and} \bibinfo{author}{\bibfnamefont{A.~G.}
  \bibnamefont{Williams}}, \bibinfo{journal}{PoS}
  \textbf{\bibinfo{volume}{LAT2007}}, \bibinfo{pages}{382}
  (\bibinfo{year}{2007}), \eprint{arXiv:0710.2410 [hep-lat]}.
%%CITATION = ARXIV:0710.2410;%%

\bibitem[{\citenamefont{Langfeld et~al.}(2002)\citenamefont{Langfeld,
  Reinhardt, and Gattnar}}]{Langfeld:2001cz}
\bibinfo{author}{\bibfnamefont{K.}~\bibnamefont{Langfeld}},
  \bibinfo{author}{\bibfnamefont{H.}~\bibnamefont{Reinhardt}},
  \bibnamefont{and} \bibinfo{author}{\bibfnamefont{J.}~\bibnamefont{Gattnar}},
  \bibinfo{journal}{Nucl. Phys.} \textbf{\bibinfo{volume}{B621}},
  \bibinfo{pages}{131} (\bibinfo{year}{2002}), \eprint{hep-ph/0107141}.
%%CITATION = HEP-PH/0107141;%%

\bibitem[{\citenamefont{Sternbeck
  et~al.}(2007{\natexlab{b}})}]{Sternbeck:2007br}
\bibinfo{author}{\bibfnamefont{A.}~\bibnamefont{Sternbeck}}
  \bibnamefont{et~al.}, \bibinfo{journal}{PoS}
  \textbf{\bibinfo{volume}{LAT2007}}, \bibinfo{pages}{256}
  (\bibinfo{year}{2007}{\natexlab{b}}), \eprint{0710.2965}.
%%CITATION = 0710.2965;%%

\bibitem[{\citenamefont{Boucaud et~al.}(2009)}]{Boucaud:2008gn}
\bibinfo{author}{\bibfnamefont{P.}~\bibnamefont{Boucaud}} \bibnamefont{et~al.},
  \bibinfo{journal}{Phys. Rev.} \textbf{\bibinfo{volume}{D79}},
  \bibinfo{pages}{014508} (\bibinfo{year}{2009}), \eprint{0811.2059}.
%%CITATION = 0811.2059;%%

\bibitem[{\citenamefont{Sternbeck et~al.}(2009)}]{Sternbeck:2010xu}
\bibinfo{author}{\bibfnamefont{A.}~\bibnamefont{Sternbeck}}
  \bibnamefont{et~al.}, \bibinfo{journal}{PoS}
  \textbf{\bibinfo{volume}{LAT2009}}, \bibinfo{pages}{210}
  (\bibinfo{year}{2009}), \eprint{1003.1585}.
%%CITATION = 1003.1585;%%

\bibitem[{\citenamefont{Langfeld}(2007)}]{Langfeld:2007zw}
\bibinfo{author}{\bibfnamefont{K.}~\bibnamefont{Langfeld}},
  \bibinfo{journal}{Phys. Rev.} \textbf{\bibinfo{volume}{D76}},
  \bibinfo{pages}{094502} (\bibinfo{year}{2007}), \eprint{0704.2635}.
%%CITATION = 0704.2635;%%

\bibitem[{\citenamefont{Bakeev et~al.}(2004)\citenamefont{Bakeev, Ilgenfritz,
  Mitrjushkin, and M{\"u}ller-Preussker}}]{Bakeev:2003rr}
\bibinfo{author}{\bibfnamefont{T.~D.} \bibnamefont{Bakeev}},
  \bibinfo{author}{\bibfnamefont{E.-M.} \bibnamefont{Ilgenfritz}},
  \bibinfo{author}{\bibfnamefont{V.~K.} \bibnamefont{Mitrjushkin}},
  \bibnamefont{and}
  \bibinfo{author}{\bibfnamefont{M.}~\bibnamefont{M{\"u}ller-Preussker}},
  \bibinfo{journal}{Phys. Rev.} \textbf{\bibinfo{volume}{D69}},
  \bibinfo{pages}{074507} (\bibinfo{year}{2004}), \eprint{hep-lat/0311041}.
%%CITATION = HEP-LAT/0311041;%%

\bibitem[{\citenamefont{Sternbeck et~al.}(2005)\citenamefont{Sternbeck,
  Ilgenfritz, M{\"u}ller-Preussker, and Schiller}}]{Sternbeck:2005tk}
\bibinfo{author}{\bibfnamefont{A.}~\bibnamefont{Sternbeck}},
  \bibinfo{author}{\bibfnamefont{E.-M.} \bibnamefont{Ilgenfritz}},
  \bibinfo{author}{\bibfnamefont{M.}~\bibnamefont{M{\"u}ller-Preussker}},
  \bibnamefont{and} \bibinfo{author}{\bibfnamefont{A.}~\bibnamefont{Schiller}},
  \bibinfo{journal}{Phys. Rev.} \textbf{\bibinfo{volume}{D72}},
  \bibinfo{pages}{014507} (\bibinfo{year}{2005}), \eprint{hep-lat/0506007}.
%%CITATION = HEP-LAT/0506007;%%

\bibitem[{\citenamefont{Cucchieri}(1997)}]{Cucchieri:1997dx}
\bibinfo{author}{\bibfnamefont{A.}~\bibnamefont{Cucchieri}},
  \bibinfo{journal}{Nucl. Phys.} \textbf{\bibinfo{volume}{B508}},
  \bibinfo{pages}{353} (\bibinfo{year}{1997}), \eprint{hep-lat/9705005}.
%%CITATION = HEP-LAT/9705005;%%

\bibitem[{\citenamefont{von Smekal et~al.}(2008{\natexlab{b}})\citenamefont{von
  Smekal, Jorkowski, Mehta, and Sternbeck}}]{vonSmekal:2008es}
\bibinfo{author}{\bibfnamefont{L.}~\bibnamefont{von Smekal}},
  \bibinfo{author}{\bibfnamefont{A.}~\bibnamefont{Jorkowski}},
  \bibinfo{author}{\bibfnamefont{D.}~\bibnamefont{Mehta}}, \bibnamefont{and}
  \bibinfo{author}{\bibfnamefont{A.}~\bibnamefont{Sternbeck}},
  \bibinfo{journal}{PoS} \textbf{\bibinfo{volume}{CONFINEMENT8}},
  \bibinfo{pages}{048} (\bibinfo{year}{2008}{\natexlab{b}}),
  \eprint{0812.2992}.
%%CITATION = 0812.2992;%%

\bibitem[{\citenamefont{Cucchieri and Mendes}(2010)}]{Cucchieri:2009zt}
\bibinfo{author}{\bibfnamefont{A.}~\bibnamefont{Cucchieri}} \bibnamefont{and}
  \bibinfo{author}{\bibfnamefont{T.}~\bibnamefont{Mendes}},
  \bibinfo{journal}{Phys. Rev.} \textbf{\bibinfo{volume}{D81}},
  \bibinfo{pages}{016005} (\bibinfo{year}{2010}), \eprint{0904.4033}.
%%CITATION = 0904.4033;%%

\bibitem[{\citenamefont{Maas}(2009)}]{Maas:2009se}
\bibinfo{author}{\bibfnamefont{A.}~\bibnamefont{Maas}} (\bibinfo{year}{2009}),
  \eprint{0907.5185}.
%%CITATION = 0907.5185;%%

\bibitem[{\citenamefont{de~Forcrand and Fromm}(2010)}]{deForcrand:2009dh}
\bibinfo{author}{\bibfnamefont{P.}~\bibnamefont{de~Forcrand}} \bibnamefont{and}
  \bibinfo{author}{\bibfnamefont{M.}~\bibnamefont{Fromm}},
  \bibinfo{journal}{Phys. Rev. Lett.} \textbf{\bibinfo{volume}{104}},
  \bibinfo{pages}{112005} (\bibinfo{year}{2010}), \eprint{0907.1915}.
%%CITATION = 0907.1915;%%

\bibitem[{\citenamefont{Davies et~al.}(1988)}]{Davies:1987vs}
\bibinfo{author}{\bibfnamefont{C.~T.~H.} \bibnamefont{Davies}}
  \bibnamefont{et~al.}, \bibinfo{journal}{Phys. Rev.}
  \textbf{\bibinfo{volume}{D37}}, \bibinfo{pages}{1581} (\bibinfo{year}{1988}).
%%CITATION = PHRVA,D37,1581;%%

\bibitem[{\citenamefont{Cucchieri and Mendes}(1998)}]{Cucchieri:1998ew}
\bibinfo{author}{\bibfnamefont{A.}~\bibnamefont{Cucchieri}} \bibnamefont{and}
  \bibinfo{author}{\bibfnamefont{T.}~\bibnamefont{Mendes}},
  \bibinfo{journal}{Phys. Rev.} \textbf{\bibinfo{volume}{D57}},
  \bibinfo{pages}{3822} (\bibinfo{year}{1998}), \eprint{hep-lat/9711047}.
%%CITATION = HEP-LAT/9711047;%%

\bibitem[{\citenamefont{Mandula and Ogilvie}(1990)}]{Mandula:1990vs}
\bibinfo{author}{\bibfnamefont{J.~E.} \bibnamefont{Mandula}} \bibnamefont{and}
  \bibinfo{author}{\bibfnamefont{M.}~\bibnamefont{Ogilvie}},
  \bibinfo{journal}{Phys. Lett.} \textbf{\bibinfo{volume}{B248}},
  \bibinfo{pages}{156} (\bibinfo{year}{1990}).
%%CITATION = PHLTA,B248,156;%%

\bibitem[{\citenamefont{Cucchieri}(1998)}]{Cucchieri:1997fy}
\bibinfo{author}{\bibfnamefont{A.}~\bibnamefont{Cucchieri}},
  \bibinfo{journal}{Phys. Lett.} \textbf{\bibinfo{volume}{B422}},
  \bibinfo{pages}{233} (\bibinfo{year}{1998}), \eprint{hep-lat/9709015}.
%%CITATION = HEP-LAT/9709015;%%

\bibitem[{\citenamefont{Leinweber et~al.}(1999)\citenamefont{Leinweber,
  Skullerud, Williams, and Parrinello}}]{Leinweber:1998uu}
\bibinfo{author}{\bibfnamefont{D.~B.} \bibnamefont{Leinweber}},
  \bibinfo{author}{\bibfnamefont{J.~I.} \bibnamefont{Skullerud}},
  \bibinfo{author}{\bibfnamefont{A.~G.} \bibnamefont{Williams}},
  \bibnamefont{and}
  \bibinfo{author}{\bibfnamefont{C.}~\bibnamefont{Parrinello}}
  (\bibinfo{collaboration}{UKQCD}), \bibinfo{journal}{Phys. Rev.}
  \textbf{\bibinfo{volume}{D60}}, \bibinfo{pages}{094507}
  (\bibinfo{year}{1999}), \eprint{hep-lat/9811027}.
%%CITATION = HEP-LAT/9811027;%%

\bibitem[{\citenamefont{Voigt et~al.}(2008)\citenamefont{Voigt, Ilgenfritz,
  M{\"u}ller-Preussker, and Sternbeck}}]{Voigt:2008rr}
\bibinfo{author}{\bibfnamefont{A.}~\bibnamefont{Voigt}},
  \bibinfo{author}{\bibfnamefont{E.-M.} \bibnamefont{Ilgenfritz}},
  \bibinfo{author}{\bibfnamefont{M.}~\bibnamefont{M{\"u}ller-Preussker}},
  \bibnamefont{and}
  \bibinfo{author}{\bibfnamefont{A.}~\bibnamefont{Sternbeck}},
  \bibinfo{journal}{Phys. Rev.} \textbf{\bibinfo{volume}{D78}},
  \bibinfo{pages}{014501} (\bibinfo{year}{2008}), \eprint{0803.2307}.
%%CITATION = 0803.2307;%%

\bibitem[{\citenamefont{Nakagawa et~al.}(2009)}]{Nakagawa:2009zf}
\bibinfo{author}{\bibfnamefont{Y.}~\bibnamefont{Nakagawa}}
  \bibnamefont{et~al.}, \bibinfo{journal}{Phys. Rev.}
  \textbf{\bibinfo{volume}{D79}}, \bibinfo{pages}{114504}
  (\bibinfo{year}{2009}), \eprint{0902.4321}.
%%CITATION = 0902.4321;%%

\bibitem[{\citenamefont{Sternbeck}(2006)}]{Sternbeck:2006rd}
\bibinfo{author}{\bibfnamefont{A.}~\bibnamefont{Sternbeck}},
  \bibinfo{journal}{PhD thesis (Humboldt University Berlin)}
  (\bibinfo{year}{2006}), \eprint{hep-lat/0609016}.
%%CITATION = HEP-LAT/0609016;%%

\end{thebibliography}

\end{document}